%% file: main_arxiv.tex
\documentclass[prl,superscriptaddress,floatfix,showpacs,notitlepage,twocolumn,longbibliography,nobibnotes
]{revtex4-2}
\usepackage{color}
\usepackage{mathtools}
\usepackage{graphicx}
\usepackage{hyperref}
\usepackage{bbold}
\usepackage{braket}
\usepackage{microtype}
\usepackage{siunitx}

\usepackage{verbatim}

\hypersetup{
  linkcolor = black,
  citecolor  = blue,
  urlcolor = blue,
  colorlinks = true,
}

\renewcommand{\vec}[1]{{\boldsymbol{#1}}}

\begin{document}
\title{Non-local mass superpositions and optical clock interferometry\\
in atomic ensemble quantum networks}

\author{Charles Fromonteil}
\affiliation{Institute for Theoretical Physics, University of Innsbruck, 6020 Innsbruck, Austria}
\affiliation{Institute for Quantum Optics and Quantum Information of the Austrian Academy of Sciences, 6020 Innsbruck, Austria}
\author{Denis V. Vasilyev}
\affiliation{Institute for Theoretical Physics, University of Innsbruck, 6020 Innsbruck, Austria}
\author{Torsten V. Zache}
\affiliation{Institute for Theoretical Physics, University of Innsbruck, 6020 Innsbruck, Austria}
\affiliation{Institute for Quantum Optics and Quantum Information of the Austrian Academy of Sciences, 6020 Innsbruck, Austria}
\author{Klemens Hammerer}
\affiliation{Institute for Theoretical Physics, University of Innsbruck, 6020 Innsbruck, Austria}
\affiliation{Institute for Quantum Optics and Quantum Information of the Austrian Academy of Sciences, 6020 Innsbruck, Austria}
\affiliation{Institut für Theoretische Physik, Leibniz Universität Hannover, Appelstraße 2, 30167 Hannover, Germany}
\author{Ana Maria Rey}
\affiliation{JILA, NIST and Department of Physics, University of Colorado, Boulder, Colorado, USA}
\affiliation{Center for Theory of Quantum Matter, University of Colorado, Boulder, Colorado, USA}
\author{Jun Ye}
\affiliation{JILA, NIST and Department of Physics, University of Colorado, Boulder, Colorado, USA}
\author{Hannes Pichler}
\email{hannes.pichler@uibk.ac.at}
\affiliation{Institute for Theoretical Physics, University of Innsbruck, 6020 Innsbruck, Austria}
\affiliation{Institute for Quantum Optics and Quantum Information of the Austrian Academy of Sciences, 6020 Innsbruck, Austria}
\author{Peter Zoller}
\email{peter.zoller@uibk.ac.at}
\affiliation{Institute for Theoretical Physics, University of Innsbruck, 6020 Innsbruck, Austria}
\affiliation{Institute for Quantum Optics and Quantum Information  of the Austrian Academy of Sciences, 6020 Innsbruck, Austria}

\begin{abstract}
Quantum networks are emerging as powerful platforms for sensing, communication, and fundamental tests of physics. We propose a programmable quantum sensing network based on entangled atomic ensembles, where optical clock qubits realize mass superpositions arising via mass-energy equivalence, as in atom and atom-clock interferometry. 
Our approach uniquely combines scalability to large atom numbers with minimal control requirements, relying only on collective addressing of internal atomic states. This enables the creation of both non-local and local superpositions with spatial separations beyond those achievable in conventional matter-wave interferometry with single atoms. 
Starting from Bell-type seed states distributed via photonic channels, collective operations within atomic ensembles coherently build many-body mass superpositions sensitive to gravitational redshift. The resulting architecture implements a non-local Ramsey interferometer, where gravitationally induced phase shifts are imprinted on non-local entangled states and are read out through local measurements at the network nodes. Beyond extending the spatial reach of mass superpositions, our scheme establishes a scalable, programmable platform to probe the interface of quantum mechanics and gravity, and offers a new experimental pathway to test atom and atom-clock interferometer proposals, e.g.~for probing gravitational dephasing, in a network-based quantum laboratory.
\end{abstract}

\maketitle

\section*{Introduction}
Understanding and probing the interplay of quantum-mechanical and gravitational effects is a major challenge of current fundamental physics research~\cite{Giulini2023Lammerzahlbook,bose2025massive,werner2024atominterferometers,chu2025exploring,Barzel2024entanglementdynamics,mieling2025quantuminterferometryexternalgravitational}.
Recent advances in the precision of atom interferometers~\cite{muller2010precisiongr,rudolph2020atomclockinterferometry,schlippert2014quantumtest,overstreet2022aharonovbohm,dobkowski2025observationquantumequivalenceprinciple,dipumpo2021redshift} and optical atomic clocks~\cite{bothwell2022redshift,zheng2023redshift,leibrandt2024ionclock,cao2024multiqubitgatesclock,marshall2025highstability} -- as well as proposed nuclear clocks~\cite{tiedau2024laserthorium,elwell2024laser,zhang2024thoriumclock,shvydko2023scandium} -- place such experiments in a unique position to explore the gravity-quantum interface. 
Given the relative weakness of gravity, one expects small general relativistic (GR) effects in such setups, dominated by a universal frequency redshift~\cite{chou2010clocksrelativity}, which has recently been measured at the millimeter scale~\cite{bothwell2022redshift,zheng2023redshift}, and can be equivalently interpreted as a result of gravitational time-dilation or of mass-energy equivalence~\cite{visib,sinha2011atominterferometry,Zych2012greffectsphotons,pikovski2017timedilation,schwartz2019postnewtonian,martinez2022abinitio}. 
Theoretical predictions range from the loss of visibility in atom-clock interferometers~\cite{visib,sinha2011atominterferometry} to so-called gravitational decoherence in spatial superpositions of macroscopic objects~\cite{decoh}. However, these have so far not been tested experimentally, as scaling up the distances, masses, and interrogation times involved is highly demanding.

To address the challenge of realizing large distances and interrogation times, we propose the use of large-scale quantum sensor networks~\cite{komar2014clocknetwork,krutyanskiy2023entanglement230m,main2025distributedqc,monroe2014largescalemodular,daiss2022gatenetwork,stolk2024metropolitan,delledonne2025operatingsystem,stas2025entanglementassistednonlocaloptical}, whose nodes consist of neutral or ionic atomic ensembles. The nodes are linked via photonic quantum channels and prepared, following established quantum repeater protocols~\cite{briegel1998repeater,duan2001longdistance,chou2005measurementinduced,liu2024metropolitannetwork,lei2023quantummemories,cao2020efficientreversible,pu2017experimentalmemory,polzik2016entanglementnetwork,hammerer2010atomensemblesreview}, in entangled states of collective excitations on long-lived transitions at the optical frequency $\omega_{eg}$. An excitation on this transition changes the internal energy by $\hbar\omega_{eg}$ and thus corresponds, via mass–energy equivalence, to an effective mass difference $m_{eg}=\hbar\omega_{eg}/c^2$. Superpositions of such excitations therefore constitute distributed mass superpositions across the network, acquiring gravitational signals equivalent to those of atom interferometers~\cite{borregaard2025testing,covey2025probing}. Using these non-local mass superpositions as seeds, we demonstrate that collective operations~\cite{pezze2018metrologyensembles,sinatra2022spinsqueezed,leroux2010cavitysqueezing,colombo2022timereversalmetrology,greve2022entanglementenhanced,li2016quantummemoryrydberg} (involving one-axis twisting) performed locally at each node can amplify the superpositions and thereby boost the phase sensitivity of the resulting sensor network states to the gravitational field. Remarkably, the resulting gravitational phase imprint can be read out by performing local measurements at each node and cross-correlating the measurement records throughout the network.

This procedure implements a generalized, non-local Ramsey interferometer enabling the observation of a range of gravitationally induced interference effects. Our scheme can reproduce the signals measured in (single-)atom interferometers without requiring any delocalization of the atoms' center of mass between the nodes. By encoding the interferometric phase in internal rather than motional degrees of freedom, it provides access to much larger effective distances and interrogation times while probing the same underlying physics through mass–energy equivalence. We illustrate the versatility of our approach using a minimal two-node network in three concrete scenarios: time dilation in states corresponding to a single-atom clock interferometer~\cite{visib}, a GR analog of the Colella–Overhauser–Werner (COW) experiment~\cite{COW}, and gravitationally induced ``decoherence"~\cite{decoh}.

The proposed scheme relies on a minimal control set, consisting only of collective light–matter interfaces, collective local control operations at each node, and local measurements. This contrasts with recent proposals~\cite{borregaard2025testing,covey2025probing} which rely  on networks of entangled single atoms or programmable quantum processors and on non-local measurements implemented via photonic interference after mapping atomic states onto light.

While our immediate goal is to realize a “gravity–quantum laboratory” for precision tests~\cite{yezoller2024essay,safronova2018newphysics,zych2018quantumeep,Mieling2022,wu2024singlephoton}, the procedures developed here also provide a versatile framework for distributed quantum metrology and other network-enabled technologies. In particular, our method of amplifying entangled seed states through optimized local operations extends previous approaches to variational optimization~\cite{cerezo2021variationalreview,kaubruegger2021variationalramsey,marciniak2022optimalmetrology}, as it involves controlling not just a single Hilbert-space vector but, at each node, an entire Hilbert subspace.

\section*{Results}

\subsection*{Gravity-quantum interference effects}

\begin{figure}
    \centering
    \includegraphics[width=\linewidth]{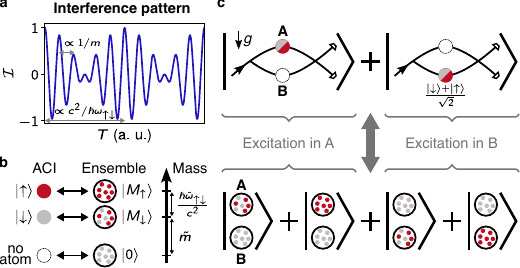}
    \caption{Example of the mapping between an ACI and a quantum network. (a) Interference pattern $\mathcal{I}(T)$ obtained by evolving a typical ACI state under gravity for a time $T$ [see Eq.~\eqref{eq:ACI_T}], and measuring it in an appropriate basis. The interference pattern shows a beat involving two distinct frequencies, determined by the atomic mass $m$ and transition frequency $\omega_{\uparrow\downarrow}$, which results in periodic decays and revivals of the signal oscillation. (b) Explicit mapping at the single-node level. The vacuum, $\ket{\downarrow}$, and $\ket{\uparrow}$ states of the interferometer arms are mapped to distinct numbers of excitations in an atomic ensemble, with masses differing by $\tilde{m}$ and $\frac{\hbar\tilde{\omega}_{\uparrow\downarrow}}{c^2}$. (c) Mapping of the full ACI state to a quantum network state. The situation corresponds to a non-local superposition of four distinct mass distributions.
    }
    \label{fig:fig1}
\end{figure}

Before turning to our main results, we briefly show that the influence of gravity on both the internal (electronic) states and the external (center-of-mass) states in a (single-)Atom Clock Interferometer (ACI) can equally well be studied with entangled states of distributed atomic ensembles, thus avoiding the need to delocalize individual atoms into spatial superpositions (see Fig.~\ref{fig:fig1}).

To be explicit, consider a single two-level atom initialized as  $\frac{1}{\sqrt{2}} \left(\ket{z_A} + \ket{z_B} \right) \otimes \frac{1}{\sqrt{2}} \left(\ket{\uparrow} + \ket{\downarrow}\right)$ in superposition of two spatial locations $z_{A/B}$ (external degree of freedom) and of its two ``clock'' states $\sigma \in \left\lbrace\uparrow, \downarrow\right\rbrace$ (internal degree of freedom)~\cite{visib}.
To leading-order, the energy $E_{z,\sigma}$ of the states $\ket{z, \sigma} = \ket{z} \otimes \ket{\sigma}$ reads
\begin{align}\label{eq:2level_atom_energies}
    E_{z,\sigma} = m c^2 + \delta_{\sigma,\uparrow} \hbar \omega_{\uparrow\downarrow} + m\phi(z) + \delta_{\sigma,\uparrow} \frac{\hbar \omega_{\uparrow\downarrow}}{c^2}\phi(z) 
\end{align}
with $m$ the atom's rest mass, $\omega_{\uparrow\downarrow}$ the energy splitting between the two internal states, and $\phi(z)$ the gravitational potential ($\delta_{\sigma,\uparrow}$ denotes the Kronecker delta function).
Crucially, in addition to the classical potential energy $m\phi(z)$, GR also predicts a gravitational redshift $\frac{\hbar \omega_{\uparrow\downarrow}}{c^2}\phi(z)$ due to the internal states.
The evolved ACI state after time $T$ is
\begin{align}\label{eq:ACI_T}
        \ket{\Psi_\text{ACI} (T)} &= \frac{1}{2} \sum_{\substack{n\in\{A,B\}\\ \sigma \in \{\uparrow, \downarrow\}}}e^{-iE_{z_n,\sigma}T/\hbar} \ket{z_n, \sigma},
\end{align}
leading to a beat between two frequencies $\hbar \, \delta\Omega = E_{z_A, \downarrow} - E_{z_B, \downarrow}  = m \left[\phi(z_A) - \phi(z_B)\right]$ and $\hbar \, (\delta\Omega + \delta\omega) = E_{z_A, \uparrow} - E_{z_B, \uparrow}  = \hbar\delta\Omega +\frac{\hbar \omega_{\uparrow\downarrow}}{c^2} \left[\phi(z_A) - \phi(z_B)\right]$
when measuring in the $\ket{\pm} = (\ket{z_A} \pm \ket{z_B})/\sqrt{2}$ basis. This results in a modulation of the visibility $\mathcal{V}=(\mathcal{I}_{\rm max}-\mathcal{I}_{\rm min})/2$ of the observed interference pattern $\mathcal{I}(T)=P_+(T)-P_-(T)$ (with $P_{\pm}$ the probability of measuring $\ket{\pm}$), as first discussed in Ref.~\cite{visib} and shown in Fig.~\ref{fig:fig1}a.
Despite tremendous progress in atom interferometers, the experimental challenges involved in scaling up the spatial separation $|z_A - z_B|$, the interrogation time $T$, and the masses in superposition, while also controlling the internal states, have so far prohibited a direct experimental observation of the predicted interference.

Instead, we make use of the fact that the same physics can also be observed with distributed excitations of internal degrees of freedom \cite{borregaard2025testing}. Specifically, we consider effective delocalized mass superpositions realized in a two-node network where each node hosts an ensemble of $N$ two-level atoms, with ground ($g$) and excited ($e$) states separated by $\hbar\omega_{eg}$, in symmetric (Dicke) states~\cite{pezze2018metrologyensembles}.
We define three states for each node: a ``massless'' state $|0\rangle$ (with respect to which masses are defined), with all atoms in $\ket{g}$, and two states $|M_\sigma\rangle$, with $\ell_\sigma$ excited atoms, corresponding to (excess) masses $M_\sigma = \ell_\sigma \, \hbar \omega_{eg}/ c^2$. Thus, Eq.~\eqref{eq:ACI_T} can be re-interpreted (see Fig.~\ref{fig:fig1}b-c) as a superposition of masses $M_\sigma$ delocalized over two locations $A/B$ corresponding to energies $E_{z_{A/B},\sigma}$, i.e.,
\begin{align}\label{eq:psi_network}
\ket{\Psi_{\text{network}}(T)}\! &=\!\frac{1}{2} \sum_{\sigma\in\{\uparrow,\downarrow\}} \left[e^{-iE_{z_A,\sigma} T/\hbar} \ket{M_\sigma}_A \otimes \ket{0}_B \right. \nonumber \\ & \left. \qquad \quad + e^{-i E_{z_B,\sigma} T/\hbar}\ket{0}_A  \otimes \ket{M_\sigma}_B \right] \;.
\end{align}
Upon identifying $m,\, \omega_{\uparrow\downarrow}$ from the ACI with 
\begin{align}
    \tilde{m} = \ell_\downarrow \frac{\hbar \omega_{eg}}{c^2} \;, && \tilde{\omega}_{\uparrow\downarrow} = \left(\ell_\uparrow - \ell_\downarrow \right) \omega_{eg} \;,
\end{align}
we see that $\ket{\Psi_{\text{network}}(T)}$ accumulates relative phases $\delta \Omega \, T$ and $\delta \omega \, T$ that exactly match those of the ACI. 
Thus, the physically observable interference pattern when performing an appropriate measurement (see below) on $\ket{\Psi_{\text{network}}(T)}$ is effectively equivalent to the one of the ACI (Fig.~\ref{fig:fig1}a).
We further elaborate on this mapping and on alternative schemes~\cite{borregaard2025testing,covey2025probing} in the Methods section.

These identifications based on the equivalence of energy and mass lie at the heart of our approach and clearly illustrate how internal states alone can be used to create the type of superpositions that occur in ACIs, and observe the corresponding gravity-quantum interference effects.
Note that this also includes an analog of the historic COW experiment by setting $\ell_\uparrow=\ell_\downarrow$. 
The original COW experiment~\cite{COW} used a neutron interferometer to observe quantum interference induced by the Newtonian gravitational potential; In our case, the effect is a consequence of GR, but leads to an equivalent interference pattern. 
These two examples (ACI and COW) also highlight the different possibilities for enhancing the experimental signal by increasing the interrogation time $T$, the involved frequency $\omega_{eg}$, or the difference in gravitational potential, essentially the distance for experiments in Earth's gravity where $\Delta\phi \equiv \phi(z_A) - \phi(z_B)  \approx g\Delta z$ with $\Delta z\equiv(z_A-z_B)$. 
As a third example, we consider so-called gravitational decoherence, as first proposed in Ref.~\cite{decoh}. In the present context, it arises from a superposition of many oscillation frequencies in a state of the form $|\Psi \rangle  = \sum_{M} \Psi_M |M\rangle_A |0\rangle_B   + (A \leftrightarrow B)$, leading to a dephasing of the resulting interference pattern. This dephasing is what we refer to as ``gravitational decoherence" in the following. 
The corresponding decoherence time is set by $\Delta E = \Delta M c^2$, where $\Delta M$ is the local mass uncertainty, 
according to (taking $\Delta\phi=g\Delta z$)~\cite{decoh}
\begin{align}\label{eq:tau_decoh}
    \tau_{\rm dec}=\frac{\sqrt2\hbar c^2}{\Delta E g \Delta z},
\end{align}
highlighting its relativistic, quantum-mechanical and gravitational origin through the presence of $c$, $\hbar$ and $g$.

In contrast to standard matter-wave interferometers, our setup relies on entanglement of the atoms' much more flexible internal states alone; In particular, clock states provide relatively large optical transition frequencies, long coherence times and---when combined with quantum communication tools---also large distances. 
Furthermore, the involved frequencies can be increased with $N$ by using more atoms per node, which is a major advantage of atomic ensembles. Another advantage of our encoding is that phase accumulation frequencies can be controlled such that oscillations and decoherence occur on the same timescale.

The examples discussed in this section---namely reduced visibility in an ACI, a GR analog of the COW experiment, and GR induced decoherence---involve special cases of non-local mass superposition states (see next section), a wide class of states which are sensitive to GR effects. 
In the following, we develop a general approach to prepare such states. For concreteness, we will focus on the three examples introduced above when discussing specific implementations. 
However, we emphasize the potential of the toolbox developed in this work to realize protocols that lie completely outside the realm of traditional interferometers, e.g., due to the constraint of particle number conservation. 
We briefly comment on further extensions in the outlook.

\subsection*{Non-local mass superpositions}

\begin{figure*}
    \centering
    \includegraphics[width=\linewidth]{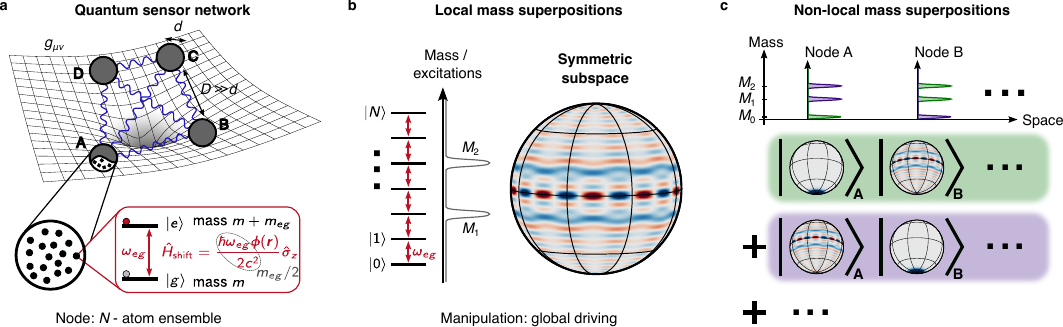}
    \caption{Non-local mass superpositions in quantum sensor networks. (a) A quantum sensor network with $N$-atom nodes evolves in a curved spacetime with metric $g_{\mu\nu}$. Each atom has two states $\ket{g}$ and $\ket{e}$, separated by an optical transition with frequency~$\omega_{eg}$, and which correspond respectively to masses $m$ and $m + m_{eg}$ (with $m_{eg}=\hbar\omega_{eg}/c^2$ the mass defect). (b) Local mass superposition in a quantum sensor node. The local Hilbert space is the symmetric subspace, spanned by a ladder of Dicke states with mass increasing by increments of~$m_{eg}$. Here we show an example of a ``clock state", i.e., a superposition of two mass eigenstates, illustrated by its Wigner function. (c) Non-local mass superpositions in the network. Several mass distribution states, including local superpositions within nodes, are superposed such that the resulting state is non-local and cannot be written as a spatial tensor product of single-node states. In this sense, non-local superpositions correspond to entanglement between the nodes. A class of states of particular interest consists in having a given local state be superposed among nodes, with the other nodes remaining in their ground states, emulating a spatial superposition of a particle with internal structure.
    }
    \label{fig:pres}
\end{figure*}

As gravity couples universally to the energy-momentum tensor of matter, its interplay with quantum mechanics can be probed with quantum superpositions of test masses. 
To this end, we aim to create and manipulate superpositions of mass distributions, i.e., states of the form
\begin{equation}\label{eq:masssup}
    \ket{\Psi}=\sum_{\vec{M}}\Psi_{\vec{M}}\ket{M}_A \otimes \ket{M'}_B \otimes \ket{M''}_C \otimes \dots \;.
\end{equation}
Here $\ket{M}_n$ are eigenstates of a mass operator $\hat{M}$ with definite mass $M$ at distinct spatial locations \mbox{$\vec{r}_n$} indexed by $n=A,B,C,\dots$. 
Eq.~\eqref{eq:masssup} thus describes a quantum superposition of several masses $\vec{M} = \left(M, M', M'', \dots\right)$ distributed in space.
We further distinguish \emph{local mass superpositions} that appear as factorized local contributions \mbox{$|\psi\rangle_A \propto |M_1\rangle_A + |M_2\rangle_A + \dots$}, from more general states that are entangled w.r.t.~the spatial tensor products in Eq.~\eqref{eq:masssup}. 
As discussed above, these latter \emph{non-local mass superpositions} clearly exhibit quantum interference phenomena that arise solely from the response of the mass superpositions to the (classical) gravitational field.

Let us make this setup explicit in a quantum sensor network consisting of several nodes $n$ distributed in space, cf. Fig.~\ref{fig:pres}a. 
We assume each node to contain an ensemble of $N$ two-level atoms with ground $(\ket{g}_i)$ and excited $(\ket{e}_i)$ states ($i=1, \dots, N$).
Specifically, we envision ensembles of atoms with a large optical energy splitting $\hbar\omega_{eg}$ that will set our unit of mass $m_{eg} = \hbar\omega_{eg}/c^2$. 
We describe gravity as a fixed classical background metric $g_{\mu\nu}$ in the limit of weak curvature, i.e., $\lvert g_{\mu\nu}-\eta_{\mu\nu}\rvert\ll 1$ where $\eta_{\mu\nu}$ is the flat Minkowski metric.
In a post-Newtonian expansion, $g_{\mu\nu}$ is well described by a gravitational potential $\phi(\vec{r})$~\cite{schwartz2019postnewtonian}. 
We further assume a separation of scales: On the one hand, individual nodes of the network are much smaller than the characteristic scale over which $\phi(\vec{r})$ varies, such that $\phi(\vec{r}_i) \approx \phi(\vec{r}_n)$ for all atoms $i$ within a node $n$ at location $\vec{r}_n$.
On the other hand, we allow for arbitrarily large distances between the nodes.

The quantum dynamics of such a network, as recorded in a suitably chosen lab frame, is governed by the Hamiltonian
\begin{align}\label{eq:Hgr}
    \hat{H}&=\sum_{\substack{n\\ i\in n}}\left[mc^2 + m\phi(\vec{r}_n) + \hbar\omega_{eg}\left(1+\frac{\phi(\vec{r}_n)}{c^2}\right)\frac{\hat{\sigma}_z^{(i)}+1}{2}\right] \nonumber\\
    &=\sum_n \left(\hat{M}_n c^2 +  \hat{M}_n \phi(\vec{r}_n) \right),
\end{align}
which can be derived as the dominant general relativistic contribution in a controlled post-Newtonian expansion in $\phi/c^2$ (see Supplementary Information)~\cite{schwartz2019postnewtonian,martinez2022abinitio,werner2024atominterferometers}. 
The Hamiltonian $\hat{H}$ generates time evolution in lab-frame coordinates ($t, \vec{r}$) where the lab is located at $\vec{r} = \vec{r}_\text{lab}$ and $\phi(\vec{r}_\mathrm{lab}) = 0$.
Here $m$ is the mass of an atom in its ground state $\ket{g}$, and the sum runs over all nodes $n$, with $i \in n$ indicating summation over all atoms $i$ within a node $n$. 
The nodes' mass operator is given by $\hat{M}_n = N m + (m_{eg}/2)\sum_{i \in n} (\hat{\sigma}_z^{(i)}+1)$.
In our setting, the main GR effect is the gravitational redshift, which couples an atom's location through the gravitational potential to its internal degrees of freedom and thus turns the mass into a dynamical quantum operator.
As a consequence, the manipulation of internal states alone allows us to explore the desired mass superpositions [Eq.~\eqref{eq:masssup}]. We emphasize that Eq.~\eqref{eq:Hgr} describes ``free'' evolution under the influence of gravity and refer to, e.g., Refs.~\cite{schwartz2019postnewtonian,martinez2022abinitio,werner2024atominterferometers} for the description of further general relativistic corrections in the presence of electromagnetic fields.

Eq.~\eqref{eq:Hgr} explicitly shows that mass eigenstates of the network are highly degenerate.
This allows us to restrict ourselves for each node to an $N+1$-dimensional local Hilbert space (see Fig.~\ref{fig:pres}b), subspace of the $2^N$-dimensional Hilbert space of the node's $N$ atoms (spanned by all computational basis states),
by selecting 
eigenstates of $\hat{M}_n$ with $N+1$ distinct eigenvalues ranging from $Nm$ to $N(m+m_{eg})$ in increments of $m_{eg}$. 
In ensembles of globally addressed neutral atoms or ions, the natural choice is the symmetric subspace, where excitations are shared among the entire ensemble: 
$\ket{0}_n=\ket{ggg\dots}_n$, $\ket{1}_n=(\ket{egg\dots}_n+\ket{geg\dots}_n+\dots)/\sqrt{N}$, \dots , $\ket{N}_n=\ket{eee\dots}_n$. 
Indeed, this subspace is accessed and preserved by collective operations.
Therefore, in the remainder of this work we will focus on the symmetric subspace (we discuss other possible subspaces in the Supplementary Information).

We can now identify the unique ground state of Eq.~\eqref{eq:Hgr} as a ``vacuum'', and collective excitations of the ensembles as various types of massive localized particles, which can be created and annihilated by locally controlling the nodes (this view can be formalized via the Schwinger representation of spin excitations, see Methods). 
Delocalizing these particles, and thereby creating non-local mass superpositions, therefore requires entangled states between several nodes (see Fig.~\ref{fig:pres}c). 
While there are many different types of entangled many-body states, let us highlight the subclass that consists of permutations $\sigma$ of single-node states, i.e., $\ket{\Psi} \sim \sum_{\sigma}\ket{\psi_{\sigma(A)}}_A\otimes\ket{\psi_{\sigma(B)}}_B\otimes\dots$, with the sum running over permutations of the nodes of the network and with local states of the form either $\ket{\psi_{\sigma(n)}}_n=\sum_{\ell=1}^{N}\psi_{\ell,\sigma(n)}\ket{\ell}_n$ (interpreted as a massive ``single-particle excitation'') or $\ket{\psi_{\sigma(n)}}_n=\ket{0}_n$ (interpreted as an ``empty'' location). The state $\ket{\Psi}$ corresponds to a spatial superposition of massive particles, with a fixed total particle number. 
In particular, a superposition in which all but one node are in the ground state~$\ket{0}_n$, while a single node---coherently delocalized over all possible positions---is in a local mass superposition~$\ket{\psi}_n$, corresponds to the states realized in standard particle interferometers, where one particle is spatially superposed across multiple locations (see Methods).

\subsection*{Preparing non-local mass superposition states}
\begin{figure}
    \centering
    \includegraphics[width=\linewidth]{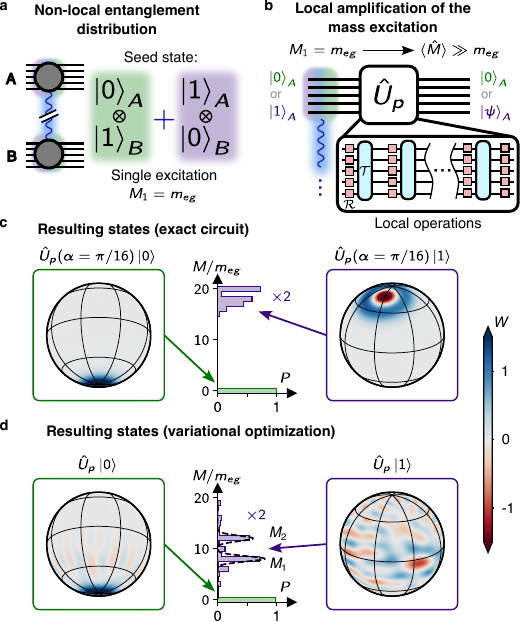}
    \caption{Preparation of non-local mass superposition states. 
    (a) Entanglement distribution between two distant nodes $A$ and $B$ generates the seed state $\ket{\Psi_0}$, in which a single mass excitation~$m_{eg}$ is delocalized across the nodes. 
    (b) Local amplification of the mass excitation. The subspace preparation unitary $\hat{\mathcal U}_p$ maps the single excitation $\ket{1}$ to the state $\ket{\psi}$, while preserving vacuum $\ket0$, producing the target state $\ket{\Psi}$ [Eq.~\eqref{eq:AB_superposition}]. $\hat{\mathcal U}_p$ can be approximated with a shallow variational circuit $\hat U_{p}$ composed of single- and multi-qubit operations, e.g., rotations and one-axis twisting. 
    (c) Action of the exact circuit $\hat{U}_p(\alpha)$, for $N=20$ atoms and $\alpha=\pi/16$ [see Eq.~\eqref{eq:exact_alpha}]. States are represented by their mass probability distributions $P(\ell)=\lvert\braket{\ell\vert\cdot}\rvert^2$ --- green for $\hat{U}_p\ket{0}$, purple (doubled) for $\hat{U}_p\ket{1}$ --- and by their Wigner distributions $W$. (d) Action of a variational circuit designed to approximate a two-mass superposition $\ket{\psi}\approx\alpha\ket{M_1}+\beta\ket{M_2}$ (illustrated by the dashed black outline; see Supplementary Information for details).}
    \label{fig:prep}
\end{figure}

We now illustrate the preparation of non-local mass superpositions for two nodes, $A$ and $B$ (see Fig.~\ref{fig:prep}).
Consider the state
\begin{align}\label{eq:AB_superposition}
\ket\Psi &= \frac{1}{\sqrt{2}} \left(\ket0_A \ket\psi_B + e^{-i\varphi_0} \ket\psi_A \ket0_B \right) \notag\\
&= \left(\hat{\mathcal U}_p^{(A)}\otimes \hat{\mathcal U}_p^{(B)}\right) \ket{\Psi_0},
\end{align}
where the local mass superposition \mbox{$|\psi\rangle_n = \sum_{\ell=1}^{N}\psi_{\ell}\ket{\ell}_n$} is delocalized across the two locations $A$ and $B$.
The second line illustrates that starting from the Bell-type seed state $\ket{\Psi_0}=\left(\ket{0}_A\ket{1}_B+e^{-i\varphi_0}\ket{1}_A\ket{0}_B\right)/\sqrt{2}$ (where we include an initial phase $\varphi_0$), the mass superposition $\ket\Psi$ can be created by applying local unitaries $\hat{\mathcal{U}}_p^{(n)}$. These local unitaries map the subspace occupied by the seed state to ${\rm span}\{\ket0_n,\ket\psi_n\}$: explicitly,
\begin{equation}\label{eq:constraint}
    \hat{\mathcal U}_p^{(n)}\ket0_{n} = \ket0_n,\quad \hat{\mathcal U}_p^{(n)}\ket1_n = \ket\psi_n\quad (n=A,B).
\end{equation}

The first step is the long-distance distribution of the seed state $\ket{\Psi_0}$ (Fig.~\ref{fig:prep}a). 
For multiple nodes, this generalizes to multipartite seeds such as W states~\cite{covey2025probing}. 
The state $\ket{\Psi_0}$ can be generated using established protocols such as quantum state transfer~\cite{cirac1997quantumstatetransfer} or heralded entanglement distribution~\cite{duan2001longdistance,sangouard2009quantumrepeaters}. 
General relativistic corrections affecting these protocols can be captured with a quantum optical model that incorporates GR effects (see Supplementary Information). 

The second step reshapes the seed by applying $\hat{\mathcal U}_p$ locally to each node $n=A,B$ (from here on, we drop the $n$ index when discussing local operations). While this transformation could be done with universal control (e.g., using quantum computers as nodes, similarly to Ref.~\cite{covey2025probing} --- see Supplementary Information and Supplementary Fig.~2), we instead adopt a more experimentally realistic setting with restricted control resources and show that the target operation $\hat{\mathcal U}_p$ can be well approximated by a low-depth variational circuit acting in the symmetric subspace (Fig.~\ref{fig:prep}b). 
Explicitly, we approximate $\hat{\mathcal U}_p$ with a variational ansatz with structure $\hat{U}_{p}(\{\vec{n}_{k},\theta_{k},\chi_{k}\}) = \prod_{k=1}^{d}\hat{\mathcal{T}}_{z}(\chi_k)\hat{\mathcal{R}}_{\vec{n}_k}(\theta_k)$ (see Supplementary Information), composed of $d$ layers of global rotations $\hat{\mathcal{R}}_{\vec{n}}(\theta)=e^{-i\theta{\vec{\hat J}}\cdot{\vec{n}}}$ and one-axis twisting (OAT) $\hat{\mathcal{T}}_{z}(\chi)=e^{-i\chi {\hat{J}_{z}}^{\,2}}$, where the collective spin operator is ${\vec{\hat J}}=\sum_{i=1}^{N}\hat{\vec{\sigma}}^{(i)}/2$ for an ensemble of $N$ atoms within a node, with $\hat{\vec{\sigma}}=\{\sigma_x,\sigma_y,\sigma_{z}\}$ the vector of Pauli operators.
These operations can be realized natively via global driving of an ensemble of neutral atoms~\cite{pezze2018metrologyensembles} or ions~\cite{marciniak2022optimalmetrology}. We then numerically optimize the variational parameters $\{\vec{n}_k,\theta_k,\chi_k\}$ using a cost function corresponding to a target subspace preparation, Eq.~\eqref{eq:constraint}. 

The complexity of this two-dimensional subspace preparation lies between variational state preparation in the symmetric sector, requiring depth $\mathcal{O}(N)$~\cite{bond2025globalvariational}, and general symmetric unitaries, which scale much less favorably. For the transformations relevant here, however, we find circuits whose depth is independent of $N$. These circuits (Fig.~\ref{fig:prep}c) are described below, whereas more general subspace transformations using low-depth circuits (Fig.~\ref{fig:prep}d) are discussed in detail in the Supplementary Information (see for instance Supplementary Fig.~1).

A particularly useful family factorizes as ${\hat{U}_p(\alpha)\equiv \hat{V}_{\alpha}\hat{U}_{\rm DT}}$. The ``double-twisting'' unitary~$\hat{U}_{\rm DT}$ (two OAT operations) creates a highly excited mass eigenstate, while~$\hat{V}_{\alpha}$ (one OAT, see Methods) tunes its energy variance. The circuit for $\hat{U}_{\rm DT}$ follows from the above variational procedure with a cost function that maximizes the energy of $\hat{U}_p\ket{1}$ while preserving $\ket0$. Remarkably, $\hat{U}_{\rm DT}$ admits an exact analytical form (see Methods).

For even atom number $N$, the two-layer circuit $\hat{U}_{\rm DT}$ maps $\ket{1}$ into the nearly maximally excited Dicke state $\ket{N-1}$ while keeping the vacuum $\ket0$ unchanged, see Fig.~\ref{fig:prep}c. Applying this operation locally to each node of a network in the seed 
state $\ket{\Psi_{0}}$ yields a highly excited non-local mass eigenstate:
\begin{equation}
\ket{\Psi_{{\rm NOON}-1}} = \frac{1}{\sqrt{2}}\left( \ket0_{A}\ket{N-1}_{B}+ e^{-i\varphi_0}\ket{N-1}_{A}\ket0_{B}\right),
\end{equation}
corresponding to the GR analog of a COW configuration with effective mass $M=(N-1)m_{eg}$~(see next section).
This state is also valuable for quantum-enhanced distributed sensing: Its quantum Fisher information~\cite{pezze2018metrologyensembles} with respect to the differential phase between the two nodes is given by $F_{Q}=(N-1)^{2}$, closely approaching the Heisenberg limit $F_{Q}=N^{2}$. The subspace preparation circuit $\hat{U}_{\rm DT}$ may therefore be useful beyond the present context.

To investigate gravitationally induced decoherence, we use $\ket{\Psi_{{\rm NOON}-1}}$ as a resource to generate a family of states featuring high mean energy and tunable energy variance. We apply $\hat{V}_{\alpha}$ locally to each node of $\ket{\Psi_{{\rm NOON}-1}}$. The resulting state of the quantum network is
\begin{align}\label{eq:exact_alpha}
\ket{\Psi_{\alpha}} &= \hat{V}_{\alpha}\otimes \hat{V}_{\alpha}\ket{\Psi_{{\rm NOON}-1}}\notag\\
&\approx\frac{ \ket0_{A}\ket{\lambda_{\alpha}}_{B}+ e^{-i\varphi_0}\ket{\lambda_{\alpha}}_{A}\ket0_{B}}{\sqrt{2}},
\end{align}
representing a spatial superposition of a tunable high-energy excitation. The operation $\hat{V}_{\alpha}$ acts as a controlled rotation: It approximately preserves the ground state $\ket0$, while rotating the excited state. The resulting state~$\ket{\lambda_{\alpha}}\approx\hat{\mathcal{R}}_{y}(2\alpha)\ket{N-1}$ has a tunable energy variance:
\begin{equation}
\frac{\Delta E^2}{\hbar^2\omega_{eg}^2} \approx\frac{(3N-2)}{4}\sin^{2}(2\alpha).
\end{equation}
This construction allows fine control over the energy variance across the range $0\leq\Delta E^2\lesssim \hbar^2\omega_{eg}^2 N$ while maintaining high mean energy---an essential feature for investigating gravitationally induced decoherence. The corresponding decoherence time $\tau_{\rm dec}$ is set by $\Delta E$ according to Eq.~\eqref{eq:tau_decoh}. The action of $\hat{U}_p(\alpha)$ on local subspace basis states is illustrated in Fig.~\ref{fig:prep}c.

Finally, we note that the entanglement distribution and subspace preparation need not be implemented in the optical qubit manifold $\{\ket{g},\ket{e}\}$. It may be experimentally more convenient to use a three-state scheme, preparing the desired entangled state in a long-lived, nearly degenerate ground state manifold $\{\ket{g},\ket{s}\}$ before turning it into a mass superposition by simultaneously applying the $\ket{s}\rightarrow\ket{e}$ transformation to all atoms, effectively ``switching on'' the gravitational redshift.

\subsection*{Non-local Ramsey interferometry with local readout}

\begin{figure}
    \centering
    \includegraphics[]{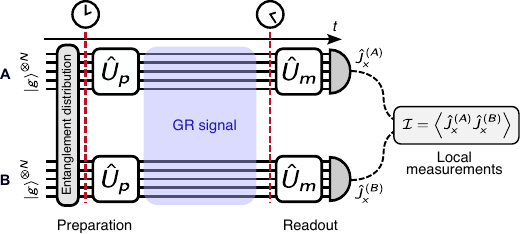}
    \caption{Non-local Ramsey interferometer with local readout. After distributing entanglement between nodes $A$ and $B$ and applying the preparation unitary $\hat{U}_p\otimes\hat{U}_p$, we let the system evolve under the Hamiltonian in Eq.~\eqref{eq:Hgr} for time $T$ (in the lab frame). We then apply the decoding unitary $\hat{U}_m\otimes\hat{U}_m$ and perform a collective $\hat{J}_x$ spin measurement at each node. The interferometric signal is given by the correlation between the two outcomes.
    }
    \label{fig:ramsey}
\end{figure}

The gravitational Hamiltonian in Eq.~\eqref{eq:Hgr} imprints mass-dependent phases on the network state. To access these phases, we introduce a non-local generalization of Ramsey interferometry with entangled input state and local measurements (Fig.~\ref{fig:ramsey}).

We begin by preparing a non-local mass superposition state as described above. The network then evolves freely under gravity for a time $T$~(defined in the laboratory frame). For the two-node state of Eq.~\eqref{eq:AB_superposition}, the evolved state reads (in a rotating frame w.r.t.~a laser that is resonant to the atomic transition $\omega_{eg}$ and located at $\vec{r}_\mathrm{lab}$)
\begin{align}\label{eq:PsiofT}
    \ket{\Psi(T)}&=\sum_{\ell=1}^N \frac{\psi_\ell}{\sqrt{2}} \left(e^{-i\varphi_{\ell,B}}\ket{0}_A\ket{\ell}_B\right. \notag\\
    &\qquad\qquad\qquad \left.+ e^{-i\varphi_0}e^{-i\varphi_{\ell,A}}\ket{\ell}_A\ket{0}_B\right),
\end{align}
where the phase $\varphi_{\ell,n}= \ell m_{eg} \phi(\vec{r}_n) T/\hbar$ depends on the excitation number $\ell$ and gravitational potential at node position $\vec{r}_n$. The gravitational redshift encoded in relative phases between different mass components can thus be extracted through an appropriate measurement on the final network state.

In contrast to earlier network-based proposals that require a non-local measurement~\cite{borregaard2025testing,covey2025probing} (e.g., mapping the quantum states of both nodes onto photonic modes and recombining the emitted fields, see Supplementary Information), we show that local measurements at each node suffice to detect gravitationally induced quantum dynamics, substantially simplifying the experimental implementation. 
Specifically, after the interrogation time $T$, we apply a local ``decoding'' unitary $\hat{U}_m = \hat{U}_p^\dagger$ at each node, followed by a collective spin measurement $\hat{J}_x$. The gravitational signal is extracted from inter-node correlations (see Methods),
\begin{equation}\label{eq:I_local}
\mathcal{I} = \langle\hat{J}_x^{(A)}\hat{J}_x^{(B)}\rangle=\frac{N}{4}\sum_{\ell, \ell'\geq 1}\lvert\psi_\ell\rvert^2 \lvert\psi_{\ell'}\rvert^2 \cos(\varphi_{\ell,B}-\varphi_{\ell',A}),
\end{equation}
which contains a superposition of oscillatory components with frequencies set by the mass-dependent phases $\varphi_{\ell,n}$, giving rise to the effects predicted in ACI experiments. The measurement variance scales as ${\sim N^2}$ for COW-like oscillations, and increases up to ${\sim N^3}$ in the regime of strong gravitational dephasing (see Methods), 
such that signal and uncertainty remain comparable for ensembles comprising hundreds to thousands of atoms. 
Note that our local measurement scheme has no analog in single-atom interferometers, due to particle number conservation (see Methods).

We emphasize that synchronization and laser phase stability are essential for the state preparation and readout steps. First, ensuring that $\hat{U}_m=\hat{U}_p^\dagger$ requires phase coherence on the scale of the Ramsey interrogation time. 
Second, both nodes must share (classical) signals such that operations are performed not only at the same lab time, but crucially also with the same laser phases. 
In this sense, what we call local operations correspond to local unitaries with classically communicated synchronization and relative control phases. 
Both requirements are addressed in state-of-the-art optical-clock~\cite{yan2025highpower} and quantum-network~\cite{irfan2025autonomousstabilizationremoteentanglement,liu2024metropolitannetwork} setups. 
We further discuss the impact of phase instability in the Supplementary Information.

\begin{figure}
    \centering
    \includegraphics[width=\linewidth]{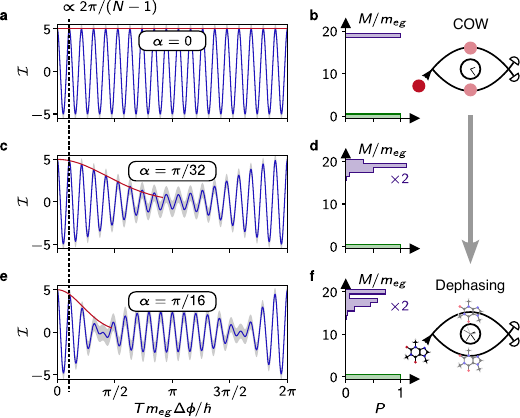}
    \caption{Interference patterns for the exact preparation circuit. (a,c,e) Interference patterns [Eq.~\eqref{eq:I_local}] for the circuit $\hat{U}_p(\alpha)$ [see Eq.~\eqref{eq:exact_alpha}], with increasing values of $\alpha$. We consider $N=20$ atoms per node and assume the $z=0$ reference clock is located at one of the nodes. The blue line represents the interferometric signal $\mathcal{I}(T)$, with the grey shaded region indicating the corresponding uncertainty band for $\kappa=100$ measurement shots (see Methods). The red line is the predicted Gaussian decay of  the interference contrast, $\exp\left[-(T/\tau_{\rm dec})^2\right]$~\cite{decoh}. Note that accounting for imperfect preparation of the seed state $\ket{\Psi_0}$ simply rescales the interference pattern $\mathcal{I}$ by the seed state fidelity (see Supplementary Information and Supplementary Fig.~3). (b,d,f) Corresponding mass probability distributions $P$ for $\hat{U}_p(\alpha)\ket{0}$ (green) and $\hat{U}_p(\alpha)\ket{1}$ (purple). The case $\alpha=0$ corresponds to a general-relativistic analog of the COW experiment~\cite{COW}, while increasing $\alpha$ leads to the emergence of gravitational dephasing~\cite{decoh}.
    }
    \label{fig:results}
\end{figure}

Fig.~\ref{fig:results} shows results of our protocol for probing gravitationally induced decoherence with $N=20$ atoms per node using preparation circuit $\hat{U}_p(\alpha)$. Resulting signals for more general low-depth circuits are discussed in the Supplementary Information (see Supplementary Fig.~1). 
Importantly, the oscillation frequencies are proportional to $N$, and in Earth’s gravitational field, the potential difference $\Delta\phi$ is proportional to the height difference: Thus, GR-induced phase accumulation can be scaled up by increasing height and particle number.

Panel~(a) displays the interference pattern for a spatial superposition of a single mass eigenvalue $M=(N-1)m_{eg}$ [see panel (b)], corresponding to $\alpha=0$. The observed oscillations occur at a frequency $\omega_{\mathcal{I}}=(N-1) \omega_{eg}\Delta \phi/c^2$, representing a relativistic analog of the classic COW experiment~\cite{COW}.
As $\alpha$ increases, the energy fluctuations grow [panels (d), (f)], leading to faster decoherence, Eq.~\eqref{eq:tau_decoh}. Panels (c), (e) show a characteristic Gaussian decay of the interference contrast, $\mathcal{I}\sim\exp[-(T/\tau_{\rm dec})^2]$, consistent with predictions from Ref.~\cite{decoh}. 
The discrepancy in panel (e) between the Gaussian envelope and the revivals observed in the signal can be explained by the bimodal mass distribution of the state [see panel (f)]. Observing these revivals provides a way to confirm the gravitational nature of the decay. 
In addition, testing the dependence of the decoherence time on energy/mass distribution of the initial state can be used to distinguish it from technical imperfections, and test whether it follows from coherent evolution with the functional dependence predicted by~GR.

\subsection*{Experimental realization scenarios}

We now discuss possible experimental implementations of the present scheme and provide quantitative estimates for near-term platforms. The non-local state-preparation and readout protocols developed above rely on two key ingredients: (i) photonic interconnects between remote nodes to distribute entanglement over large distances (in particular, Bell pairs), and (ii) globally controlled nonlinear interactions within each node (in particular OAT) to implement the local subspace preparation unitaries. Both capabilities have advanced substantially in recent years, with multiple platforms demonstrating quantum-networking functionality~\cite{krutyanskiy2023entanglement230m,liu2024metropolitannetwork} as well as collective unitary control~\cite{pogorelov2021compactiontrap,bohnet2016quantumspin,bornet2023scalable,eckner2023squeezing,franke2023quantumenhanced}.

Among the platforms that meet our requirements, trapped-ion quantum-network architectures are particularly promising. Local nodes can be realized as programmable trapped-ion systems equipped with optical-cavity-based light-matter interfaces. Recent experiments have demonstrated heralded remote entanglement generation over distances of up to \(\qty{230}{\meter}\) with fidelities of \(88\%\)~\cite{krutyanskiy2023entanglement230m}, as well as OAT with sufficiently large nonlinearity to generate highly entangled states, including GHZ states of up to 24 ions~\cite{pogorelov2021compactiontrap}.
Motivated by these capabilities, we first provide quantitative resource estimates for a trapped-ion implementation.

We consider an ensemble of \(N=20\) (globally addressed) ions per node, with mass superposition states encoded in the \(\mathrm{^{40}Ca^+}\) optical clock transition at \(729\,\mathrm{nm}\) (i.e., \(\omega_{eg}=2\pi\times 411\,\mathrm{THz}\)), with lifetime \(1/\Gamma\approx 1\,\mathrm{s}\). The central requirement for observing GR-induced dynamics is that the coherence time \(\tau_{\max}\) exceed the characteristic timescale of the effect we want to see. There are three such timescales: to see a single oscillation, \(\tau_{\mathrm{osc}} \sim 2\pi c^{2}/(N\omega_{eg} g\,\Delta z)\); to see gravitational dephasing, $\tau_{\rm dec} \sim 2\sqrt{2}c^2/(\sqrt{{3N}}\omega_{eg}g\Delta z)$ (for $\alpha$ chosen to minimize the dephasing time); and to see full revivals of the oscillations, $\tau_{\rm rev} \sim 2\pi c^2/(\omega_{eg}g\Delta z)$. For the NOON-like states considered here, spontaneous decay typically limits the coherence time to \(\tau_{\max}\sim 1/(N\Gamma)\). Thus, the requirement that $\tau_{\rm max}$ be much larger than each of these timescales translates, respectively: to $\Delta z\gg 2\pi \Gamma c^2/(\omega_{eg} g)\sim 22\,\rm m$ for a single oscillation; to $\Delta z\gg 2\sqrt{2N/3}\Gamma c^2/(\omega_{eg} g)\sim 26\,\rm m$ for gravitational dephasing; and to $\Delta z\gg 2\pi N\Gamma c^2/(\omega_{eg} g)\sim 450\,\rm m$ for the revivals. The first two conditions lie within the capabilities of state-of-the-art network experiments~\cite{krutyanskiy2023entanglement230m}, while the third one will require further improvements. Increasing \(\Delta z\) further reduces the total interrogation time required to resolve the interference pattern.

To resolve the interference fringes, the signal must be much larger than its statistical uncertainty, i.e., \(\mathcal{I}\gg\Delta \mathcal{I}/\sqrt{\kappa}\) for $\kappa$ measurement shots (this discussion can also be adapted to the case of estimating gravitational phases, see Supplementary Information). Since the single-shot uncertainty typically scales as \(\Delta \mathcal{I}\sim \sqrt{2}\,\mathcal{I}\) (see Methods), this is satisfied with, e.g., \(\kappa= 200\) shots per time point. With 100 points, this corresponds for \(\Delta z=50\,\mathrm{m}\) to a total interrogation time \(T_{\mathrm{tot}}\sim 100\,\kappa\,\tau_{\mathrm{dec}}/2 \sim 260\,\mathrm{s}\). The distance and time requirements could be further relaxed by employing ion species with longer excited-state lifetimes. In summary, given ongoing progress towards entanglement distribution over distances of tens of kilometres, observation of the effects discussed in the present work appears feasible with near-term trapped-ion network technology. We note that the above derivations can also be carried out for other (in particular, neutral-atom) platforms, giving similar results.

Beyond OAT, scaling mass-superposition states to large particle numbers is naturally achievable in programmable quantum simulators with global-control capabilities and finite-range interactions, e.g., in trapped ions and Rydberg-tweezer arrays. In both trapped ions and neutral-atom arrays in 1D and 2D geometries, effective spin models with tunable long-range interactions of the form \(\sum_{i<j}{\sigma_i^{z}\sigma_j^{z}}/{r_{ij}^{\alpha}}\) can be implemented. For trapped ions, the exponent \(\alpha\) is broadly tunable, \(0<\alpha<3\), whereas Rydberg-mediated interactions typically yield \(\alpha=6\) (or \(\alpha=3\), depending on the dressing scheme). This finite-range, power-law behavior contrasts with the effectively infinite-range coupling implicit in OAT, which we used above as a resource for generating mass superpositions. A complementary architecture is provided by atomic ensembles coupled to high-cooperativity optical cavities, which combine photonic network connectivity with cavity-mediated long-range interactions~\cite{barberena2024tradeoffs}. While substantial spin squeezing has been demonstrated in both trapped-ion and neutral-atom setups~\cite{bornet2023scalable,eckner2023squeezing,franke2023quantumenhanced}, these platforms also support variational protocols based solely on global control~\cite{marciniak2022optimalmetrology}. In this setting, low-depth variational circuits provide a practical route to engineer large-\(N\) mass-superposition states using experimentally accessible operations. Moreover, the underlying computationally hard task of preparing and manipulating many-body states in higher-dimensional geometries with finite-range interactions can be tackled via on-device optimisation, as demonstrated in trapped ions in the context of variational Ramsey interferometry~\cite{marciniak2022optimalmetrology}. Adapting these ideas suggests a clear pathway for implementing analogous variational quantum protocols within the network-based non-local Ramsey interferometer proposed here.

Lastly, increasing both the transition energy and the excited-state lifetime would further enhance the scheme. In this context, recent progress toward nuclear clocks with ultraviolet transition frequencies~\cite{tiedau2024laserthorium,elwell2024laser,zhang2024thoriumclock} and even X-ray transitions~\cite{shvydko2023scandium} is particularly noteworthy. Incorporating such systems into our proposed setup will require further advances in nuclear-clock technology, in particular coherent control and the ability to prepare entangled states of nuclear qubits.

\section*{Discussion}
In this work, we have shown how to make use of a network of atomic ensembles in order to put so-far elusive general relativistic effects on quantum systems within reach of current experimental capabilities.
While our focus lies on tests of fundamental physics, we expect that the toolbox laid out here will translate to other scenarios such as distributed quantum sensing~\cite{proctor2018multiparameter,eldredge2018optimal,ge2018distributed,gessner2020multiparameter,sekatski2020optimal,pezze2025distributed,guo2020distributed,malia2022distributed,bate2025experimental}, e.g., of electromagnetic fields.

Although our illustrations have focused on a minimal network consisting of two nodes, the approach can be generalized to multiple nodes. 
This would enable, e.g., a direct measurement of all components of the Riemann curvature tensor, which has only been partially achieved so far~\cite{rosi2015measurement,asenbaum2017phaseshiftcurvature}. 
It will be interesting to construct network states that are tailored to the measurement of the Riemann tensor or other quantities derived from the gravitational field like the Ricci and Kretschmann scalars.

Finally, it will be interesting to extend our approach from quasi-static to dynamic spacetimes as relevant for, e.g., gravitational wave detection~\cite{cahillane2022review,graham2013gwatomic,khalili2018overcoming}.
Here, more detailed studies are necessary to identify which kind of states entangled across large distances could provide improved sensitivity to gravitational waves in frequency bands less accessible with traditional interferometers~\cite{aggarwal2025challenges}.

\small 

\section*{Methods}
\subsection*{Mapping to atom-clock interferometers}
Here we give more details on our identification of networked atomic ensembles with atom interferometers. 

Consider again the superposition of a single two-level atom in two spatial locations $z_{A/B}$ and its internal states $\sigma \in \left\lbrace\uparrow, \downarrow\right\rbrace$,
\begin{align}\label{eq:ACI_state}
    \ket{\Psi_\text{ACI}}  &= \frac{1}{\sqrt{2}} \left(\ket{z_A} + \ket{z_B} \right) \otimes \frac{1}{\sqrt{2}} \left(\ket{\uparrow} + \ket{\downarrow}\right)\notag \\
    &= \frac{1}{2} \left(\hat{\psi}^\dagger_{z_A, \uparrow} + \hat{\psi}^\dagger_{z_A, \downarrow} + \hat{\psi}^\dagger_{z_B, \uparrow} + \hat{\psi}^\dagger_{z_B, \downarrow}\right)  \ket{\text{vac.}} \;.
\end{align}
Here, we expressed the state in second-quantized notation with operators $\hat{\psi}^\dagger_{z,\sigma}$ creating a particle of spin $\sigma$ and mass $m$ at location $z$ from the vacuum $\ket{\text{vac.}}$, highlighting the structure of the state as a single excitation in superposition. 

With the interpretation of mass superpositions in mind, consider again two atomic ensembles at two fixed locations $z_{n}$ with $n\in\left\lbrace A,B\right\rbrace$, where we describe the symmetrized state of each ensemble with two internal states $\alpha \in \{g,e\}$ per atom, via the Jordan-Schwinger representation~\cite{pezze2018metrologyensembles}, using bosonic operators $\hat{a}^{\dagger}_{\alpha}$ and $\hat{b}^{\dagger}_{\alpha}$ creating a particle at $A$ and $B$, respectively.  
In direct correspondence with Eq.~\eqref{eq:ACI_state}, we construct an entangled network state as
\begin{align}
        \ket{\Psi_\text{network}}  &= \frac{1}{\sqrt{2}}\left[\frac{1}{\sqrt{2}} \left(\ket{\ell_\uparrow}_A + \ket{\ell_\downarrow}_A\right)  \otimes \ket{0}_B +  \left( A \leftrightarrow B \right)\right]\\
        &= \frac{1}{2} \left[ \mathcal{N}(\ell_\uparrow) \left(\hat{a}_{e}^\dagger\right)^{\ell_\uparrow} \left(\hat{a}_{g}^\dagger\right)^{N-\ell_\uparrow} \left(\hat{b}_{g}^\dagger\right)^{N} \right. \nonumber\\
        &\qquad +\left. \mathcal{N}(\ell_\downarrow) \left(\hat{a}_{e}^\dagger\right)^{\ell_\downarrow} \left(\hat{a}_{g}^\dagger\right)^{N-\ell_\downarrow} \left(\hat{b}_{g}^\dagger\right)^{N} \right.\nonumber\\
        &\qquad + \left( a \leftrightarrow b \right) \Big] \ket{\text{vac.}}_A \otimes \ket{\text{vac.}}_B\;,
\end{align}
where $\ket{\ell}_{n} =  \frac{\left(\hat{c}_e^\dagger \right)^{\ell}  \left(\hat{c}_g^\dagger\right)^{N-\ell}}{\sqrt{\ell !} \sqrt{(N-\ell)!}} \ket{\text{vac.}}_{n}$ with $\hat{c}=\hat{a}(\hat{b})$ for $n=A(B)$ denotes an $\ell$-excitation state of a single ensemble, $\ket{\text{vac.}}_{n}$ is the local node state without particles, and we abbreviated the normalization factors $\mathcal{N}(\ell) = 1/\sqrt{\ell ! (N-\ell)! N!}$.
While the first line above partially obscures the direct correspondence between $\ket{\Psi_\text{network}}$ and $\ket{\Psi_\text{ACI}}$, because the tensor product distinguishes the two spatial nodes of the network rather than the external vs. internal states as in Eq.~\eqref{eq:ACI_state},
the second-quantized notation more clearly identifies a single collective excitation in superposition of $\uparrow$ and $\downarrow$ delocalized over~$A$ and~$B$. 

Taking the atoms in the network to have mass $m$ and energy splitting $\omega_{eg}$, with energy given by $E_{z,g}$ and $E_{z,e}$ [in analogy to Eq.~\eqref{eq:2level_atom_energies}], the corresponding energies of the four states involved in the superposition are given by
\begin{align}
    \tilde{E}^{(N)}_{z_A,\sigma} &= \ell_\sigma E_{z_A,e} + (N-\ell_\sigma) E_{z_A,g} + N E_{z_B,g} \nonumber\\
    &= \ell_\sigma \left(E_{z_A, e} - E_{z_A, g}\right) + \text{const} \;, \\
    \tilde{E}^{(N)}_{z_B,\sigma} &= N E_{z_A,g} +  \ell_\sigma E_{z_B,e} + (N-\ell_\sigma) E_{z_B,g} \nonumber\\&= \ell_\sigma \left(E_{z_B, e} - E_{z_B, g}\right) + \text{const}  \;,
\end{align}
which leads to the identifications stated in the main text. 
In particular, this justifies identifying $\ket{\ell_{\sigma}}_n \leftrightarrow \ket{M_{\sigma}}_n$ as a mass eigenstate with $M_\sigma = \ell_\sigma \left(E_{z_n,e} - E_{z_n,g}\right)/c^2$ that depends on the location $n$ of the node. 
Note that the assignment of zero mass to the no-excitation state $\ket{0}_n$ corresponds to measuring energy w.r.t.~$\text{const} = N(E_{z_A,g} + E_{z_B,g})$. Dropping the constant introduces no observable effects within our framework for fixed particle number $N$ and fixed locations $z_{A/B}$ of the nodes.

We note that the identification presented here is not the only way to map the ACI scheme to a quantum network setup. 
For instance, Ref.~\cite{borregaard2025testing} considers an example of two entangled qutrits, with three levels $\ket{g}$, $\ket{a}$ and $\ket{b}$ identified as the vacuum, $\ket{\downarrow}$ and $\ket{\uparrow}$ states of the interferometer, and related work~\cite{covey2025probing} 
extends this to multiple network nodes, specifically proposing a class of W-like network states of atomic processors.
In contrast to this, 
our ensemble-based approach emphasizes scalability: It avoids single-atom control, leverages efficient manipulation of large atomic ensembles, and naturally generalizes to more complex states---particularly those relevant for exploring
gravitational decoherence~\cite{decoh}.

\subsection*{Measurements in the non-local Ramsey interferometer}
Here we detail the derivation of the interference signal $\mathcal{I}$ in Eq.~\eqref{eq:I_local} for the local measurement scheme of the main text (see Fig.~\ref{fig:ramsey}), as a function of the gravitational phases $\varphi_{\ell,n}$; we also give expressions for the uncertainty $\Delta \mathcal{I}$ of this measurement; in addition, we discuss its relationship with the measurement that naturally occurs in an atom interferometer. In this section we take $\varphi_0=0$ for simplicity, but extension to the case of a non-vanishing $\varphi_0$ is straightforward. 

In an atom interferometer, the final measurement is understood to be performed on the position degree of freedom, i.e., regardless of the internal state. This corresponds in our framework to the observable:
\begin{align}\label{eq:O_if}
    \hat{\mathcal O}_{\rm at.\, int.}=\sum_{\ell\geq1} \ket{\Psi_{\ell,+}}\bra{\Psi_{\ell,+}}-\ket{\Psi_{\ell,-}}\bra{\Psi_{\ell,-}},
\end{align}
where we have defined $\ket{\Psi_{\ell,\pm}}=(\ket{0}_A\ket{\ell}_B\pm\ket{\ell}_A\ket{0}_B)/\sqrt{2}$. 
Applying this measurement to the final state in Eq.~\eqref{eq:PsiofT} leads to the following expectation value:
\begin{align}
    \mathcal{I}_{\rm at.\, int.}(T)&=\left\langle\hat{\mathcal O}_{\rm at.\, int.}\right\rangle=\sum_{\ell\geq 1}\lvert\psi_\ell\rvert^2\cos(\varphi_{\ell,B}-\varphi_{\ell,A}).\label{eq:positionmeasurement}
\end{align}
Let us now compute the analogous interference signals in the local measurement scheme defined in the main text.

We first apply a decoding unitary consisting in undoing the excitation amplification step ($\hat{U}_m=\hat{U}_p^{\dagger}$). Then, we measure the collective spin operators $\hat{J}_x^{(A)}$ and $\hat{J}_x^{(B)}$ simultaneously. Finally, we take as signal the (classically computed) product of the two outcomes, i.e., $\mathcal{I}(T)=\langle \hat{J}_x^{(A)}\hat{J}_x^{(B)} \rangle$. The idea behind this measurement is the following: In the case of a single mass eigenstate $\ket{\ell}$ in superposition, the decoding step brings the network back to a single-excitation state $\hat{U}_m\ket{\Psi(T)}=e^{-i\varphi_{\ell,B}}\ket{0}_A\ket{1}_B+e^{-i\varphi_{\ell,A}}\ket{1}_A\ket{0}_B$, which is a Bell-type state with a differential phase. In a two-qubit system, this Bell-state phase can be measured by a coordinated $\hat{\sigma}_x\hat{\sigma}_x$ measurement, without requiring an entangled measurement of the two qubits. In our case, the $\hat{\sigma}_x$ operators are replaced by collective spin-$x$ operators.
This measurement scheme has no analog in conventional (single-) atom interferometers. Indeed, the local $\hat{J}_x$ observables correspond in that setting to quadrature-like operators: For instance, in the COW case (i.e, superposed particle with no internal structure), the operator is $\hat{\psi}_{z_{A/B}}+\hat{\psi}_{z_{A/B}}^\dagger$ [using the notation from Eq.~\eqref{eq:ACI_state}]. These are not accessible local observables, as they do not conserve particle number.

In detail, the measurement observable (including the decoding step) reads $\hat{\mathcal O}_{\rm loc}=(\hat{U}_m^{\dagger} \otimes \hat{U}_m^{\dagger})\hat{J}_x^{(A)}\hat{J}_x^{(B)} (\hat{U}_m \otimes \hat{U}_m)$. 
Moreover, it can be shown that for any $\ell,\ell'\geq 1$ and $\sigma,\sigma'=\pm1$,
\begin{align}\label{eq:Jxa_Jxb_elements}
    \bra{\Psi_{\ell,\sigma}}\hat{J}_x^{(A)}\hat{J}_x^{(B)}\ket{\Psi_{\ell',\sigma'}}= \frac{N}{4}\sigma\, \delta_{\ell 1}\delta_{\ell\ell'}\delta_{\sigma\sigma'}
\end{align}
[with $\ket{\Psi_{\ell,\pm}}$ defined above, see Eq.~\eqref{eq:O_if}].
Since the $\ket{\Psi_{\ell,\sigma}}$'s form a basis of the space of possible $\ket{\Psi(T)}$ (which is stable under $\hat{U}_m\otimes \hat{U}_m$), the projection of the observable $\hat{\mathcal O}_{\rm loc}$ on this space can be written as
\begin{align}
    \hat{\mathcal O}_{\rm loc}&=\sum_{\substack{\ell,\ell'\geq 1\\\sigma,\sigma'=\pm 1}}(\hat{U}_m^{\dagger} \otimes \hat{U}_m^{\dagger})\ket{\Psi_{\ell,\sigma}}\notag\\
    &\qquad\qquad\qquad\bra{\Psi_{\ell,\sigma}}\hat{J}_x^{(A)}\hat{J}_x^{(B)} \ket{\Psi_{\ell',\sigma'}}\bra{\Psi_{\ell',\sigma'}}(\hat{U}_m \otimes \hat{U}_m)\notag\\
    &=\frac{N}{4}\sum_{\sigma=\pm1}(\hat{U}_m^{\dagger} \otimes \hat{U}_m^{\dagger})\ket{\Psi_{1,\sigma}}\sigma\bra{\Psi_{1,\sigma}}(\hat{U}_m \otimes \hat{U}_m).
\end{align}
Additionally, we have $(\hat{U}_m^{\dagger} \otimes \hat{U}_m^{\dagger})\ket{\Psi_{1,\pm}}=\ket{\Psi_{\pm}}$ where \begin{align}
\ket{\Psi_{\pm}}\equiv\sum_{\ell\geq 1}\frac{\psi_\ell}{\sqrt{2}}(\ket{0}_A\ket{\ell}_B \pm \ket{\ell}_A\ket{0}_B).
\end{align}
Thus, measuring $\hat{\mathcal O}_{\rm loc}$ for the final state $\ket{\Psi(T)}$ in Eq.~\eqref{eq:PsiofT} gives expectation value
\begin{align}
    \mathcal{I}(T)&=\frac{N}{4}\sum_{\sigma=\pm1}\sigma\left\lvert\braket{\Psi(T) | \Psi_{\sigma}}\right\rvert^2\notag\\
    &=\frac{N}{4}\sum_{\ell,\ell'\geq 1}\lvert\psi_\ell\rvert^2 \lvert\psi_{\ell'}\rvert^2 \cos(\varphi_{\ell,B}-\varphi_{\ell',A}),\label{eq:signal_localmeasurement}
\end{align}
which is the result in Eq.~\eqref{eq:I_local}. We note that for a single mass eigenstate $\ell$, we obtain $\mathcal{I}(T)=\frac{N}{4}\cos(\varphi_{\ell,B}-\varphi_{\ell,A})$, which is the expected COW-like signal oscillation. 
We further note that this scheme requires a shared laser phase (both to ensure $\hat{U}_m=\hat{U}_p^{\dagger}$ for both nodes and to ensure that the same spin component is measured for both nodes). The choice of this laser is tied to that of the lab frame. In particular, for a reference frame (i.e., reference clock position) chosen such that either $\varphi_{\ell,B}$ or $\varphi_{\ell,A}$ vanishes, we obtain $\mathcal{I}(T)=\frac{N}{4}\sum_{\ell}\lvert\psi_\ell\rvert^2  \cos(\varphi_{\ell,B}-\varphi_{\ell,A})$, reproducing the result of Eq.~\eqref{eq:positionmeasurement} up to a prefactor. Thus, the local measurement scheme has the same qualitative properties as the position measurement, with the practical advantage of not requiring any faithful transport of quantum information between the nodes. 

We can now similarly derive the uncertainty associated to our measurement scheme, $\Delta\mathcal{I}^2=\left\langle{\hat{\mathcal{O}}_{\rm loc}}^2\right\rangle - \left\langle\hat{\mathcal{O}}_{\rm loc}\right\rangle^2.$
The first term can be expressed as 
\begin{align}
    \langle{\hat{\mathcal{O}}_{\rm loc}}^2\rangle&=\bra{\Psi(T)}{\hat{\mathcal{O}}_{\rm loc}}^2\ket{\Psi(T)}\notag\\
    &=\sum_{\substack{\ell,\ell'\geq 1\\ \sigma,\sigma'=\pm 1}}\lambda_{\ell,\sigma}\bra{\Psi_{\ell,\sigma}}(\hat{J}_x^{(A)}\otimes \hat{J}_x^{(B)})^2\ket{\Psi_{\ell',\sigma'}}\lambda_{\ell',\sigma'}^{*},
\end{align}
where we defined $\lambda_{\ell,\sigma}\equiv\bra{\Psi(T)}(\hat{U}_p\otimes \hat{U}_p)\ket{\Psi_{\ell,\sigma}}$.
Since the $\hat{J}_x$ operators are now squared, $\bra{\Psi_{\ell,\sigma}}(\hat{J}_x^{(A)}\otimes \hat{J}_x^{(B)})^2\ket{\Psi_{\ell',\sigma'}}$ is nonzero also for $\ell,\ell'\neq 1$. Computing the expression of $\bra{\Psi_{\ell,\sigma}}(\hat{J}_x^{(A)}\otimes \hat{J}_x^{(B)})^2\ket{\Psi_{\ell',\sigma'}}$ finally gives
\begin{align}\label{eq:Oloc_2}
    \langle{\hat{\mathcal{O}}_{\rm loc}}^2\rangle&=\frac{1}{16}\sum_{\ell,\sigma}\left\lvert\lambda_{\ell,\sigma}\right\rvert^2 \left[(2\ell+1)N^2-2\ell^2N+\delta_{\ell 2} N(2N-2)\right]
\end{align}
This expression scales typically as $N^2$, and in the worst case as $N^3$. Taking for instance $T=0$, we have $\bra{\Psi(T)}(\hat{U}_p\otimes \hat{U}_p)=\bra{\Psi_{1+}}$, which gives us that
\begin{align}
    \Delta\mathcal{I}({T=0})&=\frac{\sqrt{N(N-1)}}{2\sqrt{2}}.
\end{align}
Comparing this to the expectation value $\mathcal{I}(T=0)=N/4$ gives $\Delta\mathcal{I}\sim\sqrt{2}\mathcal{I}$, which is what we use for the numerical estimates of the main text. The effect of laser phase noise on the signal uncertainty $\Delta\mathcal{I}$, as well as the impact of $\Delta\mathcal{I}$ on phase estimation sensitivity, are discussed in detail in the Supplementary Information.

Lastly, we note that the measurement scheme described in the main text is not the only way to observe gravitationally-induced interference; in the Supplementary Information, we discuss alternative possibilities.

\subsection*{Preparation of non-local mass superpositions: Exact circuit}

Here we provide explicit constructions for the unitaries $\hat{U}_{\rm DT}$ and $\hat{V}_{\alpha}$. Together, these operations enable the generation of highly excited non-local cat states with tunable mean energy and energy variance, as discussed in the main text.

The double-twisting unitary, $\hat{U}_{\rm DT}$, is composed of two consecutive OAT operations about orthogonal axes:
\begin{align}
\hat{U}_{\rm DT}\equiv\hat{\mathcal{T}}_{y}\left(\pm\frac{\pi}{2}\right)\hat{\mathcal{T}}_{x}\left(\frac{\pi}{2}\right),
\end{align}
where the sign of the second twisting angle depends on the parity of the total spin $S=N/2$: positive for odd $S$, negative for even. If experimental constraints restrict implementation to positive twisting angles only, the unitary can be implemented for even $S$ as $\hat{U}_{\rm DT}=\hat{\mathcal{R}}_{y}(\pi)\hat{\mathcal{T}}_{y}\left(\frac{\pi}{2}\right)\hat{\mathcal{T}}_{x}\left(\frac{\pi}{2}\right)$.

For even atom number $N$, the action of $\hat{U}_{\rm DT}$ can be represented in the permutationally symmetric subspace spanned by the Dicke states $\ket{\ell}\equiv\ket{S,-S+\ell}$, with $\ell=0,\ldots,N$, as:
\begin{align}
\hat{U}_{\rm DT} &= e^{\frac{i\pi}{4} \left((-1)^S-1\right)}\sum_{\ell\in\mathrm{even}}\ket{\ell}\bra{\ell}\notag\\
&+ e^{\frac{i\pi}{4} \left(3-(-1)^S\right)}\sum_{\ell\in\mathrm{odd}}\ket{N-\ell}\bra{\ell}.
\end{align}
This decomposition shows that $\hat{U}_{\rm DT}$ preserves the components with even $\ell$ while mapping the odd-$\ell$ components to $\ket{N-\ell}$---effectively performing a state-controlled reflection in the $\hat{J}_z$ eigenbasis.

The local unitary $\hat{V}_{\alpha}$ is designed to act as a conditional rotation which preserves the vacuum $\hat{V}_{\alpha}\ket0 \approx\ket0$ while rotating the excited state $\hat{V}_{\alpha}\ket{N-1} =\ket{\lambda_{\alpha}}\approx\hat{\mathcal{R}}_{y}(2\alpha)\ket{N-1}$. It is defined as:
\begin{align}
\hat{V}_{\alpha}=\hat{\mathcal{R}}_{f}\,\hat{\mathcal{T}}_{z}\left(\chi\right)\hat{\mathcal{R}}_{y}(\alpha),
\end{align}
where the rotation angle $\alpha$ controls the resulting energy variance, and the twisting angle is chosen as: 
\begin{equation}
\label{eq:chi_for_V}
\chi=\frac{(1+2k)\pi}{2(2S-1)\cos\alpha},\; k\in\mathbb{Z}.
\end{equation}
The final rotation $\hat{\mathcal{R}}_{f}$ is selected to preserve the vacuum state, i.e., it maximizes $\bra0 \hat{V}_{\alpha}\ket0$.

To understand the action of the nonlinear OAT gate $\hat{\mathcal{T}}_{z}\left(\chi\right)$
on the states $\hat{\mathcal{R}}_{y}(\alpha)\ket0$ and $\hat{\mathcal{R}}_{y}(\alpha)\ket{N-1}$,
we use a mean-field approximation. For the twisting Hamiltonian $\hat{H}_{\mathcal T}=\kappa \hat{J}_{z}^{2}$,
the equations of motion for the spin expectation values are:
\begin{align}
\partial_t{\braket{\hat{J}_{x}}} & =i\braket{[\hat{H}_{\mathcal T},\hat{J}_{x}]}=-\kappa\braket{\hat{J}_{y}\hat{J}_{z}+\hat{J}_{z}\hat{J}_{y}}\approx-2\kappa\braket{\hat{J}_{z}}\braket{\hat{J}_{y}},\notag\\
\partial_t{\braket{\hat{J}_{y}}} & \approx2\kappa\braket{\hat{J}_{z}}\braket{\hat{J}_{x}},\\
\partial_t{\braket{\hat{J}_{z}}} & =0.\notag
\end{align}
This indicates that the twisting operation effectively implements
a state-dependent rotation about the $z$-axis with angular frequency
$2\kappa\braket{\hat{J}_{z}}$, leading to the approximation:
\begin{align}
\hat{\mathcal{T}}_{z}\left(\chi\right)\approx\hat{\mathcal{R}}_{z}\left(2\chi\braket{\hat{J}_{z}}\right).
\end{align}
This leads to an approximate form for $\hat{\mathcal{R}}_f$ that restores the vacuum component:
\begin{align}
\hat{\mathcal{R}}_{f}\approx \hat{\mathcal{R}}_{y}(-\alpha)\hat{\mathcal{R}}_{z}\left(2\chi S\cos\alpha\right),
\end{align}
since $\braket{\hat{J}_{z}}=-S\cos\alpha$ for the state $\hat{\mathcal{R}}_{y}(\alpha)\ket0$.

We now evaluate the action of $\hat{V}_{\alpha}$ on the excited Dicke state $\ket{N-1}=\ket{S,S-1}$, which has $\braket{\hat{J}_{z}}=(S-1)\cos\alpha$ after rotation. Using the same approximation for the OAT gate, we find:
\begin{align}
\hat{V}_{\alpha}\ket{N-1} & =\hat{\mathcal{R}}_{f}\hat{\mathcal{T}}_{z}\left(\chi\right)\hat{\mathcal{R}}_{y}(\alpha)\ket{N-1}\notag\\
 & \approx\hat{\mathcal{R}}_{y}(-\alpha)\hat{\mathcal{R}}_{z}\left(2\chi S\cos\alpha\right)\notag\\
 &\quad\times\hat{\mathcal{R}}_{z}\left(2\chi(S-1)\cos\alpha\right)\hat{\mathcal{R}}_{y}(\alpha)\ket{N-1}\notag\\
 & =\hat{\mathcal{R}}_{y}(-\alpha)\hat{\mathcal{R}}_{z}\left(2\chi(2S-1)\cos\alpha\right)\hat{\mathcal{R}}_{y}(\alpha)\ket{N-1}.
\end{align}
Choosing the twisting angle according to Eq.~\eqref{eq:chi_for_V} ensures that the net rotation becomes $\hat{\mathcal{R}}_{y}(2\alpha)$, up to a global phase. This confirms that $\hat{V}_{\alpha}$ acts as a conditional rotation, affecting only the excited component of a non-local cat state.

\section*{Data availability}
The data generated in this work have been deposited in the Zenodo digital repository under \url{https://doi.org/10.5281/zenodo.22029217}.

\vspace{-0.5cm}
\section*{Code availability}
The code used in this work is available from the corresponding author upon request.

\bibliography{bibl}

\vspace{-0.5cm}
\section*{Acknowledgements}
The authors thank Shraddha Agrawal, Johannes Borregaard, Časlav Brukner, Wolfgang Dür, Christopher Hilweg, Raphael Kaubrügger, Ben Lanyon, Igor Pikovski, and Philip Walther for valuable discussions and comments on the manuscript. C.F., T.V.Z., and P.Z. gratefully acknowledge the hospitality of JILA (University of Colorado Boulder) during research visits supported by Q-SEnSE.

\vspace{-0.5cm}
\section*{Funding}
This work is part of the QuantA (Quantum Austria) initiative, funded
by the Austrian Science Fund (FWF) [Grant DOI: 10.55776/COE1]. 
It was further supported by the European Research Council (ERC) under the European Union’s Horizon 2020 research and innovation programme (grant agreement No. 101041435, ERC Starting Grant QARA to Hannes Pichler), and by the European Union’s Horizon Europe research and innovation programme (grant agreement No. 101113690, PASQuanS2). Work in Boulder was supported by the U.S. National Science Foundation through its Quantum Leap Challenge Institute program, via Q-SEnSE (Quantum Systems through Entangled Science and Engineering, NSF Award OMA-2016244) at the University of Colorado Boulder, by the Sloan Foundation, the Simons Foundation and the Heising-Simons Foundation, and NSF JILA-PFC PHY-2317149.

\section*{Author Contributions}
C.F., D.V.V., T.V.Z., and P.Z. conceived the project, developed the theoretical framework, and prepared the initial draft of the manuscript. C.F., D.V.V., T.V.Z., K.H., A.M.R., J.Y., H.P., and P.Z. contributed to discussions, refinement of the concepts, and revision of the final manuscript.

\section*{Competing interests}
The authors declare no competing interests.

\include{SI_arxiv}

\end{document}

%% file: SI_arxiv.tex
\clearpage

\onecolumngrid
\begin{center}
    {\Large\bfseries Supplementary Information for ``Non-local mass superpositions and optical clock interferometry in atomic ensemble quantum networks"}
\end{center}
\vspace{2em}
\twocolumngrid

\normalsize

\renewcommand{\theequation}{S\arabic{equation}}
\renewcommand{\figurename}{Supplementary Fig.}
\renewcommand{\thetable}{S\arabic{table}}

\setcounter{equation}{0}
\setcounter{figure}{0}
\setcounter{table}{0}

\section{General relativistic corrections to Hamiltonian dynamics}
We now discuss the derivation of the relativistically-corrected system Hamiltonian as a power series in $c^{-2}$, as given in Eq.~(7) of the main text. While we refer the reader to the literature~\cite{schwartz2019postnewtonian,martinez2022abinitio,werner2024atominterferometers} for a detailed derivation of the corrections, we provide here a condensed discussion for completeness.

The spacetime metric, in the post-Newtonian formalism, can be expressed via the line element
\begin{align}
    ds^2=-\left(1+2\frac{\phi(\vec{r})}{c^2} +2\frac{\phi(\vec{r})^2}{c^4}\right)c^2 dt^2+\left(1-2\frac{\phi(\vec{r})}{c^2}\right)d\vec{r}^2,
\end{align}
in an appropriate coordinate system and with $\phi(\vec{r})$ the shifted gravitational potential such that $\phi(\vec{r}_{\rm lab})=0$~\cite{werner2024atominterferometers}. 

By considering atoms as composite particles, and writing their Lagrangian using the above metric, including the atom-light interaction term, the leading-order corrections to the dynamics can be derived \cite{schwartz2019postnewtonian}, such that an atom's Hamiltonian can be written as
\begin{align}
    \hat{H}=\hat{H}_M+\hat{H}_I+\hat{H}_{M-I}+\hat{H}_{A-L}+O\left(c^{-4}\right),
\end{align}
where $\hat{H}_M$ is the center-of-mass Hamiltonian of the atom, $\hat{H}_I$ the internal one, 
$\hat{H}_{M-I}$ the relativistic coupling between these two degrees of freedom, and $\hat{H}_{A-L}$ the atom-light coupling Hamiltonian.

In our setting, we focus solely on the internal degree of freedom of the atoms. As $c^{-2}$-order corrections to the center-of-mass d.o.f. influence the internal one via $\hat{H}_{M-I}$, itself of order $c^{-2}$, they can be ignored as they lead to higher-order effects. The correction terms in $\hat{H}_I$ are absorbed in the definition of states $\ket{g}$ and $\ket{e}$. As a consequence, for ``free" evolution (without light) we need only consider the correction terms in $\hat{H}_{M-I}$, which read
\begin{align}
    \hat{H}_{M-I}&=\left[\phi(\hat{\vec{r}})-\frac{\hat{\vec{p}}^2}{2m^2}\right]\otimes\frac{\hat{H}_I}{c^2}+\hat{H}_{\rm metric},\label{eq:Hmi}
\end{align}
where $\hat{\vec{r}}$ and $\hat{\vec{p}}$ are the atom's position and momentum operators. 
$\hat{H}_{\rm metric}$ is a metric correction term which is fully off-diagonal and can be neglected in the rotating wave approximation~\cite{werner2024atominterferometers,martinez2022abinitio}.
For cold atomic ensembles, the contribution from $\vec{\hat{p}}^2$ (second-order Doppler shift) can be neglected with respect to the gravitational redshift (e.g., for an ensemble of $^{87}\rm Sr$ atoms at 25~{\textmu}K and for a network distance of $\qty{1}{\meter}$, the resulting fractional frequency shifts are respectively $\sim4\times10^{-20}$ and $\sim10^{-16}$). When it is not negligible, the second-order Doppler shift results in a systematic error term which can nevertheless be accounted for if the temperature is known. In addition, in our setup the atoms are localized such that we take the center-of-mass degrees of freedom to be classical variables $\hat{\vec{r}}\rightarrow\vec{r}_n$. 
Thus, the sum of the non-relativistic Hamiltonian and the corrections leads to the Hamiltonian $\hat{H}$ in Eq.~(7) of the main text.

The atom-light interaction term plays a role during state preparation. Its effect can be analyzed by writing the system Hamiltonian as a quantum optical model $\hat{H}=\hat{H}_A+\hat{H}_L+\hat{H}_{A-L}$, where $\hat{H}_A=\hat{H}_I + \hat{H}_{M-I}$ is the atomic Hamiltonian (summed over all atoms), $\hat{H}_L=\int d^3\vec{k}\;\hbar\omega \hat{b}^{\dagger}(\omega)\hat{b}(\omega)$ is the EM field Hamiltonian (with relativistically corrected modes \cite{werner2024atominterferometers}), and $\hat{H}_{A-L}$ has been rewritten as
\begin{align}\label{eq:Hal}
\hat{H}_{A-L}&=\int  d^3\vec{k}\;\left[\frac{\kappa(\omega)}{2}\hat{b}^{\dagger}(\omega)\hat{\sigma}_- e^{i\Phi_{\vec{k}}(\vec{r})}+{\rm h.c.}\right],
\end{align}
with $\Phi_{\vec{k}}(\vec{r})=(1-\phi(\vec{r})/c^2)\vec{k}\cdot\vec{r}$ the relativistically-corrected phase. From this, a quantum state transfer protocol between two nodes can be described via a quantum stochastic Schr\"odinger equation, which shows the three main corrections from the non-relativistic case to be (i) a systematic phase shift (which can be absorbed in $\varphi_0$) (ii) a correction to the propagation time between the two nodes, and (iii) a detuning term proportional to $\hat{\sigma}_z/c^2$, which leads to a small state preparation infidelity. Corrections (i) and (ii) stem from the relativistic correction to the electromagnetic field modes in $\hat{H}_{A-L}$ [Eq.~\eqref{eq:Hal}], while correction (iii) corresponds to the redshift term in $\hat{H}_{M-I}$ [Eq.~\eqref{eq:Hmi}]. As the phase shift leads to a systematic change of $\varphi_0$, and the state preparation is much shorter than the interrogation time, we neglect these corrections in the analysis of the main text.

The choice of rotating frame we made in the main text can be understood from this model. Considering a single-mode laser with frequency $\omega_L$ in a coherent state, we replace $\hat{b}(\omega)$ with $\beta e^{-i\omega_L t}$, such that $\hat{H}_L$ becomes constant, and $\hat{H}_{A-L}$ becomes $\hat{H}_{A-L}=\frac{\kappa}{2}\beta^* e^{i\omega_L t} e^{i\Phi_\vec{k}(\vec{r})}\hat{\sigma}_- + \rm h.c.$. Then, we go to a frame rotating with the laser frequency and take the laser to be resonant with the atoms at $\vec{r}=\vec{r}_{\rm lab}$, i.e., $\omega_L=\omega_{eg}$: the corresponding transformation is $\hat{U}_{\rm rot}=e^{i\omega_{eg}t\ket{e}\bra{e}}$, eventually leading to the replacements $\hat{H}_A\rightarrow\tfrac{\hbar\omega_{eg}\phi(\vec{r})}{c^2}\ket{e}\bra{e}$ and $\hat{H}_{A-L}\rightarrow\tfrac{\kappa}{2}\beta^* e^{i\Phi_{\vec{k}}(\vec{r})}\hat{\sigma}_- + \rm h.c.$.

\section{Details on variational circuits}
We now detail how circuits realizing the local preparation unitary $\hat{\mathcal{U}}_p$ can be designed through variational optimization of the circuit parameters.

The resources we consider are global rotations ($\hat{\mathcal{R}}_{\vec{n}}(\theta)$) and OAT ($\hat{\mathcal{T}}_{\vec{n}}(\chi)$) in the symmetric subspace, as described in the main text. We optimize the parameters of a variational circuit consisting of layers of $\hat{\mathcal{R}}$ and $\hat{\mathcal{T}}$, with the variational ansatz $\hat{U}_{p}(\{\vec{n}_{k},\theta_{k},\chi_{k}\})$ given in the main text. 
For our numerical optimization, we eliminate redundancies in the parameterization by rewriting the ansatz as:
\begin{align}
    \hat{U}_p(\{\vec{x}_{k}\})=e^{-i \vec{\hat{J}}\cdot\vec{x}_{d+1}}\prod_{k=1}^{d} e^{-i (\vec{\hat{J}}\cdot\vec{x}_{k})^2}
\end{align}
where $\vec{x}_k=(x_k,y_k,z_k)$
; this corresponds to $d$ layers of (arbitrary) OAT and one final rotation. This is equivalent to $\hat{U}_{p}(\{\vec{n}_{k},\theta_{k},\chi_{k}\})$ (specifically, with $\chi_k=\|\vec{x}_k\|^2$), since rotations between the OAT operations can be absorbed in the $\vec{x}_k$ parameters. This ansatz has $3d+3$ parameters, which can be reduced to $3d+1$ due to the $z$-rotation symmetry of the problem (specifically, by setting $x_1=z_{d+1}=0$).

We consider two types of cost functions, depending on the state preparation goal: to approximate a given mass superposition state, or to optimize the properties of the resulting state. 
For approximating a given target state, the cost function has two requirements: First, it should encode the constraint $\hat{U}_p\ket{0}=\ket{0}$; Second, it should quantify the overlap of the prepared excitation $\hat{U}_p\ket{1}$ with the target state $\ket{\psi}$. In fact, only the overlap between the mass probability distributions of $\hat{U}_p\ket{1}$ and $\ket{\psi}$ are necessary, as relative phases between different excitation numbers in $\ket\psi$ are irrelevant to the Ramsey interferometry results discussed in the main text [see Eq.~(14)]. Thus, the cost function in this case is
\begin{align}\label{eq:cost_state}
    \mathcal{C}(\{\vec{x}_k\})=-\lvert\bra{0}\hat{U}_p\ket{0}\rvert^2 - \lambda\sum_{\ell \geq 1}\lvert\braket{\ell\vert\psi}\rvert\,\lvert\bra{\ell}\hat{U}_p\ket{1}\rvert.
\end{align}
with $\lambda$ a hyperparameter determining the relative importance of vacuum and target fidelities.

Another option for the choice of cost function is to optimize properties of the prepared state. For instance, if we want to observe gravitational oscillations and decoherence but not to probe a specific state, a natural goal is to maximize the mean value and variance of the prepared mass excitation. 
An example of cost function fitting this goal is
\begin{align}\label{eq:cost_nrj_var}
    \mathcal{C}(\{\vec{x}_k\})=-\lvert\bra{0}\hat{U}_p\ket{0}\rvert^2 - \lambda_1\frac{\langle\hat{J}_z\rangle}{N} - \lambda_2\frac{\sqrt{\langle\hat{J}_z^2\rangle-\langle\hat{J}_z\rangle^2}}{N},
\end{align}
where the expectation values $\langle \cdot \rangle$ are all evaluated on the state $\hat{U}_p\ket{1}$. The parameters $\lambda_1$ and $\lambda_2$ can be chosen to control the relative weight of the expectation value and variance. In particular, by choosing $\lambda_2=0$ we can retrieve the double-twisting unitary $\hat{U}_{\rm DT}$.

\begin{figure}
    \centering
    \includegraphics[width=\linewidth]{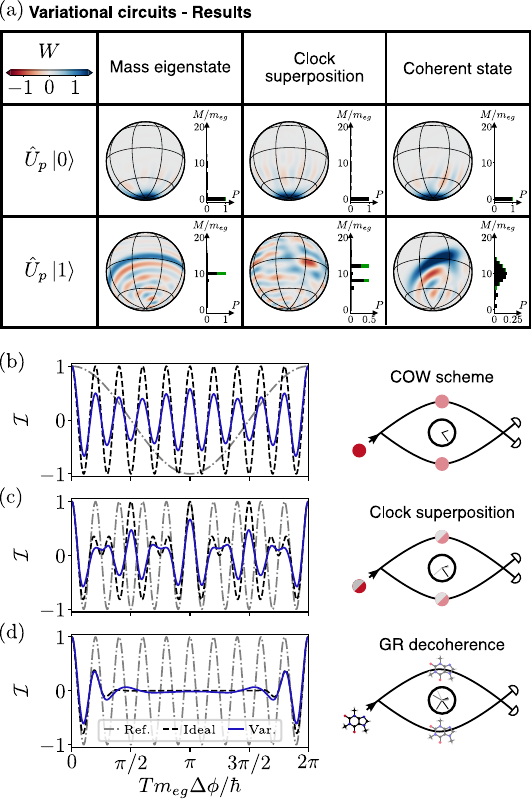}
    \caption{Results for variational circuits. (a) Resulting states $\hat{U}_p\ket{0}$ and $\hat{U}_p\ket{1}$ for variationally optimized circuits, with $N=20$. Results are shown for three distinct target excitations: single mass eigenstate, clock superposition, and coherent state. States are illustrated by their Wigner distributions $W$ and their mass probability distributions (in black, with target distributions in green). (b-d) Resulting interference patterns [with the measurement scheme of Eq.~\eqref{eq:I_nonlocal}], for ideal (black) and variational (blue) states. The reference lines (grey) are COW-like interference patterns for mass $m_{eg}$ [in (b)] or for the states' average mass [in (c-d)]. (b) Single mass eigenstate, reproducing the COW experiment~\cite{COW,muller2010precisiongr}. (c) Superposition of two mass eigenstates, corresponding to the clock superposition proposal of Ref.~\cite{visib}. (d) Coherent spin state, illustrating gravitationally induced decoherence~\cite{decoh}.}
    \label{fig:varia}
\end{figure}

The typical circuits we obtain from variational optimization with Eq.~\eqref{eq:cost_state} can be intuitively interpreted as follows: A first, large-angle OAT layer ($\chi\sim\pi/2$) separates the $\ket{0}$ and $\ket{1}$ states into two distinct locations on the Bloch sphere, followed by small OAT layers with a well-chosen axis to further push these two states apart. Finally, a global rotation brings $\ket{0}$ back to the south pole of the Bloch sphere to ensure $\hat{U}_p\ket{0}=\ket{0}$. This naturally generates excitations $\ket{\psi}$ whose Wigner function looks like a continuous ``cloud" on the Bloch sphere. A more complex state such as the clock superposition requires two additional large-twisting layers, similar to the $\hat{U}_{\rm DT}$ unitary of the main text, to go from a single eigenstate to a superposition.

Supplementary Fig.~\ref{fig:varia} shows resulting states and corresponding interference patterns (for $N=20$) for variational circuits optimized for preparing three example target excitations [via Eq.~\eqref{eq:cost_state}]. These correspond to the three example cases discussed in the main text:

$(1)$ A single mass eigenstate $\ket{\psi} = \ket{M}$. The variational circuit consists of three OAT layers, with parameters 
\begin{align}
\{\vec{x}_k\}&=\{(0,0.25,1.27),(0.19,0.13,-0.45),\notag\\
&\qquad(0.32,-0.14,-0.35),(0.29,0.42,0)\}.
\end{align}
This case is an equivalent of the classic COW experiment~\cite{COW}, with a signal oscillating at a single frequency $M \Delta\phi/\hbar$, with $\Delta\phi$ the difference of gravitational potential between the two nodes.
We attribute the loss of visibility to deviations of the target state from a perfect mass eigenstate superposition.

$(2)$ A ``clock'', i.e., a superposition of two mass eigenstates, $\ket{\psi} = (\ket{M_1} + \ket{M_2})/\sqrt{2}$. The circuit consists of five OAT layers: a three-layer circuit $\hat{U}_p(\{\vec{x}_k\})$ with parameters 
\begin{align}
\{\vec{x}_k\}&=\{(0,-0.24,1.26),(-0.03,-0.18,-0.53),\notag\\
&\qquad(-0.02,0.11,-0.54),(0,03,-0.08,0)\},
\end{align}
followed by two layers $\mathcal{T}_y(\pi/2)\mathcal{R}_z(2.04)\mathcal{T}_y(\pi/2)$.
This case is equivalent to an atom interferometry scheme with a two-level atom in an internal state superposition~\cite{visib}. 
It thus leads to a beat between two distinct frequencies, which can be observed as periodic losses and revivals of the signal oscillation.
Once again, this is faithfully reproduced by our variational state preparation.

$(3)$ A coherent spin state $\ket{\psi} = \hat{\mathcal{R}}_y(\pi/2)\ket{0}$~\cite{pezze2018metrologyensembles}. The variational circuit consists of three OAT layers, with parameters \begin{align}
\{\vec{x}_k\}&=\{(0,-0.22,1.25),(0.18,-0.09,-0.70),\notag\\
&\qquad(-0.04,-0.20,-0.29),(1.03,-0.21,0)\}.
\end{align} This is an example of a more complex internal state, where a local superposition of many eigenstates is delocalized over the two nodes. 
This evidences the effect of gravitational decoherence~\cite{decoh} as the interference of accumulated frequencies leads to dephasing and a fast loss of oscillations. 

Overall, these results demonstrate that relatively shallow circuits are indeed capable of preparing states with mass distributions close to the target ones, and that the resulting interference patterns display the expected features.

\section{Alternative choices of states}\label{sec:alt_states}
\subsubsection{Beyond the symmetric subspace}
In the main text, we have focused on excitations in the symmetric subspace of the atomic ensembles, as this is the natural subspace explored with collective control. However, what truly matters is that we have access to $N+1$ basis states with masses differing by increments of $m_{eg}$. If the entangled seed state is outside of the symmetric subspace, collective local operations during the state amplification step will also generate states outside of the symmetric subspace.

This is particularly relevant for near-term implementations: in current trapped-ion-network setups, entanglement is typically distributed among a pair of ions (say $a$ for node $A$ and $b$ for node $B$), instead of the initial excitation being delocalized over an entire ensemble. 

To be specific, let us consider such a seed state. It can be written as
\begin{align}
    \ket{\Psi_0}=\ket{\mathcal{B}_+}_{ab} \otimes \ket{g}^{\otimes N-1}_{A\setminus a} \otimes \ket{g}^{\otimes N-1}_{B\setminus b},
\end{align}

where $a$ (resp. $b$) is the first qubit of node $A$ (resp. $B$), and $\ket{\mathcal{B}_+}_{ab}=(\ket{g}_a\ket{e}_b+\ket{e}_a\ket{g}_b)/\sqrt{2}$ is a Bell state between $a$ and $b$.

Applying the OAT-based circuits of the main text to this seed state generates mass superpositions analogous to those of the main text, but which live in a different subspace of the Hilbert space. For instance, the unitary $\hat{U}_{\rm DT}$ preserves $\ket{g}^{\otimes N}$ and sends $\ket{e}\otimes\ket{g}^{\otimes N-1}$ to $\ket{g}\otimes\ket{e}^{\otimes N-1}$, which is an $N-1$-excitation mass eigenstate.

The resulting mass superposition state $(\hat{U}_p\otimes\hat{U}_p)\ket{\Psi_0}$ can then be interrogated for a Ramsey time $T$, and, with an adapted measurement scheme (see Sec.~\ref{sec:alt_measurements} below), produces the same interference signals as in the main text.

\subsubsection{Individual addressing}

Our scheme can also be extended to the case of individually controlled optical qubits; in that case, the excitation amplification step can be performed using controlled gates. This variant breaks the global control requirement of the main text, hindering scalability; however, it may be of interest to see signatures of gravitational decoherence in small-scale systems.

We choose an $N+1$-dimensional subspace spanned by a basis in which the added excitations are on specific qubits: $\ket{0}=\ket{g}^{\otimes N}$, $\ket{1}=\ket{egg\dots}$, $\ket{2}=\ket{eeg\dots}$, etc. until $\ket{N}=\ket{e}^{\otimes N}$. Classic entanglement distribution protocols can generate the Bell-like seed state $\ket{\Psi_0}$ in this case by simply creating a shared Bell pair between the first qubits of the two nodes. Then, the local preparation circuit $\hat{U}_p$ can be expressed in terms of controlled unitaries on the first qubit (see Supplementary Fig.~\ref{fig:circuit_ions}): The control ensures that the vacuum state $\ket{0}$ remains untouched, while the excitation $\ket{1}$ changes. For example (identifying $\ket{g}$ and $\ket{e}$ with the $0$ and $1$ states of a qubit, respectively), the circuit 
\begin{align}
    \hat{U}_p={\rm CNOT}_{1\ell}\dots{\rm CNOT}_{13}{\rm CNOT}_{12}
\end{align}
(where ${\rm CNOT}_{ij}$ denotes a controlled-NOT gate on qubits $i$ and $j$) corresponds to the target state $\ket{\psi}=\ket{\ell}$ (single mass eigenstate), while the circuit
\begin{align}
    \hat{U}_p={\rm CNOT}_{\ell_1\ell_2}\dots{\rm CNOT}_{\ell_1\ell_1+1}{\rm CH}_{1\ell_1}\dots{\rm CNOT}_{13}{\rm CNOT}_{12}
\end{align}
(where ${\rm CH}_{ij}$ is a controlled-Hadamard gate on qubits $i$ and $j$) prepares the target state $\ket{\psi}=(\ket{\ell_2}+\ket{\ell_1-1})/\sqrt{2}$ (two-eigenstate mass superposition).

Due to the seed state being different in this case from the one of the main text, the measurement scheme also needs to be slightly adapted, see Sec.~\ref{sec:alt_measurements}.

\begin{figure}
    \centering
    \includegraphics[width=\linewidth]{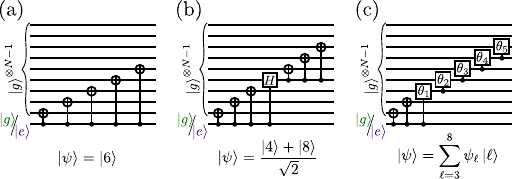}
    \caption{Preparation of three example target states in an individually-controlled variant of our scheme. Quantum gates are controlled on the first qubit of the node being in $\ket{e}$. The $H$ square denotes a Hadamard gate, while the $\theta$ ones denote qubit rotations of angle $\theta$ around the y-axis. The three target states are: (a) A single mass eigenstate (with six excitations), prepared using a series of $\rm CNOT$ gates. (b) A superposition of two mass eigenstates (with four and eight excitations), prepared with $\rm CNOT$ gates and a controlled-Hadamard. (c) A superposition of several eigenstates (from three to eight excitations), prepared with $\rm CNOT$ gates and controlled-rotations. The probabilities are given by $\lvert\psi_\ell\rvert^2 = \cos^2(\theta_{\ell-2}/2)\prod_{\ell'=3}^\ell \sin^2(\theta_{\ell'-3}/2)$. For instance, the choice $\vec{\theta}/\pi=(.79,.71,.63,.54,.42)$ leads to a Gaussian-like probability distribution comparable to the coherent spin state of the main text.}
    \label{fig:circuit_ions}
\end{figure}

\section{Impact of experimental imperfections}

In the main text, our results have focused on ideal entanglement distribution and squeezing operations; in addition, we have assumed perfect laser phase stability (in both space and time) and no decay during the interrogation time. Here we discuss the impact of these imperfections on our scheme.
\paragraph{Imperfect entanglement distribution.} To study the impact of imperfect entanglement distribution, we use a toy model assuming that errors are dominated by dark counts. That is, we replace the Bell-type seed state $\ket{\Psi_0}$ with a mixed state of the form:
\begin{align}\label{eq:imp_entanglement}
    \hat{\rho}_0(\mathcal{F}_{\rm ent})=\mathcal{F}_{\rm ent}\ket{\Psi_0}\bra{\Psi_0}+(1-\mathcal{F}_{\rm ent})\ket{0_A0_B}\bra{0_A0_B};
\end{align}
i.e., the correct state is prepared with probability $\mathcal{F}_{\rm ent}$ (which is the fidelity of the entanglement distribution), and otherwise no excitations are present in the network.

We can then reproduce the computation of the interferometric signal from the main text, with $\hat{\rho}_0$ as initial state. Both terms of the mixture evolve independently, and $\ket{0_A0_B}$ is preserved by the preparation unitary (by design) and by the GR-corrected evolution (since $\varphi_{0,A}=\varphi_{0,B}=0$). Thus, the final state $\hat{\rho}_T(\mathcal{F}_{\rm ent})$ is a mixture of $\ket{\Psi(T)}$ and $\ket{0_A0_B}$. Finally, the interferometric signal is given by
\begin{align}
    \mathcal{I}(\mathcal{F}_{\rm ent})&={\rm Tr}\left[\hat{\rho}_T(\mathcal{F}_{\rm ent}) \hat{\mathcal{O}}_{\rm loc}\right]\\
    &=\mathcal{F}_{\rm ent} \bra{\Psi(T)}\hat{\mathcal{O}}_{\rm loc}\ket{\Psi(T)}\notag\\
    &\qquad+(1-\mathcal{F}_{\rm ent}) \bra{0_A0_B}\hat{\mathcal{O}}_{\rm loc}\ket{0_A0_B}.
\end{align}
The state $\ket{0_A0_B}$ is preserved by the decoding unitary, and a $\hat{J}_x$ measurement on the state $\ket{0}$ has a vanishing expectation value. Thus, the second term vanishes, and the interference pattern $\mathcal{I}$ is simply the ideal one rescaled by a factor $\mathcal{F}_{\rm ent}$. We note that the above analysis can be generalized beyond considering only dark counts: For instance, including localized single-excitations ($\ket{0_A1_B}$, $\ket{1_A0_B}$) leads to the same result, since $\ket{0}$ is preserved by the preparation and decoding.

\paragraph{Imperfect one-axis twisting.} We now consider imperfections in the state amplification step, i.e., imperfect implementation of OAT. As an example, for trapped-ion OAT gates (see main text), GHZ state preparation has been demonstrated with order 99\% fidelities for few ions, and for up to 24 ions with 54\% fidelity~\cite{pogorelov2021compactiontrap}.

Similar to the case of imperfect entanglement, we study this with a simplified toy model, replacing the preparation unitary $\hat{U}_p$ with a quantum channel which applies $\hat{U}_p$ with probability $\mathcal{F}_{\rm OAT}$ and returns a maximally mixed state with probability $1-\mathcal{F}_{\rm OAT}$, i.e., locally:
\begin{align}
    \Lambda_{p}[\rho]=\mathcal{F}_{\rm OAT}\hat{U}_p\rho \hat{U}_p^{\dagger} + (1-\mathcal{F}_{\rm OAT})\mathbb{1}/d
\end{align}
(with $d$ the local Hilbert space dimension). The decoding channel has a similar form, replacing $\hat{U}_p$ with $\hat{U}_p^{\dagger}$. 

After the state preparation and Ramsey interrogation steps, the state is $\mathcal{F}_{\rm OAT}^2 \ket{\Psi(T)}\bra{\Psi(T)}$ plus terms involving the maximally mixed state $\mathbb{1}$ at at least one of the nodes. Similarly, after decoding, we end up with $\mathcal{F}_{\rm OAT}^4 (\hat{U}_p^{\dagger}\otimes\hat{U}_p^{\dagger})\ket{\Psi(T)}\bra{\Psi(T)}(\hat{U}_p\otimes\hat{U}_p)$ plus terms involving the maximally mixed state. Finally, we perform a joint measurement of the spin-$x$ components, $\hat{J}_x^{(A)}\otimes\hat{J}_x^{(B)}$. On a maximally mixed state, the expectation value of the measurement is ${\rm Tr}[\hat{J}_x/d]=0$; thus, the contributions to the interference signal $\mathcal{I}=\left\langle \hat{J}_x^{(A)}\otimes\hat{J}_x^{(B)}\right\rangle$ from all terms other than the first one vanish. Based on this toy model, we conclude that imperfect one-axis twisting primarily rescales the interferometric signal by a factor $\mathcal{F}_{\rm OAT}^4$. Therefore, the lower the fidelity, the more measurement shots will be needed to resolve the signal. Note that $\mathcal{F}_{\rm OAT}$ will depend on the particle number $N$~\cite{pogorelov2021compactiontrap}.

\paragraph{Phase instability.} Our scheme has two main phase stability requirements: stability in space (i.e., operations are performed with the same phase at both nodes) and stability in time (i.e., the state preparation and readout steps are performed with the same phase). We discuss these two cases separately below.

We model phase instability in time by adding a random phase $\varepsilon$ (we take for simplicity a Gaussian probability distribution $p(\varepsilon)$ with vanishing expectation value, and variance $\sigma_\varphi^2$) to the decoding unitary $\hat{U}_m$ and to the collective spin-measurement operator $\hat{J}_x$ for both nodes. This description can be derived from the complete analysis of the frequency-noise power spectral density~\cite{yan2025highpower}; we consider the $\sigma_\varphi\ll1$ limit, and derive the leading-order corrections in $\sigma_\varphi$ to the interferometric signal $\mathcal{I}$ and its uncertainty $\Delta\mathcal{I}$. Using experimental data from Ref.~\cite{yan2025highpower}, we estimate that $\sigma_\varphi^2\sim 4\times 10^{-5}$ for $25 \, \rm ms$ of interrogation time (consistent with the numerical estimates of the main text). 
The instability causes our measurement observable $\hat{\mathcal{O}}_{\rm loc}$ (see Methods section) to become $\hat{\mathcal{O}}_{\rm loc}(\varepsilon)=(e^{i\varepsilon \hat{J}_z}\otimes e^{i\varepsilon \hat{J}_z})\hat{\mathcal{O}}_{\rm loc}(e^{-i\varepsilon \hat{J}_z}\otimes e^{-i\varepsilon \hat{J}_z})$. Adapting the derivation of the Methods section, we obtain the following expression for the interferometric signal, i.e., the expectation value for measuring $\hat{\mathcal{O}}_{\rm loc}(\varepsilon)$ on the final (pure) state $\ket{\Psi(T)}$:
\begin{align}
    \mathcal{I}&=\int p(\varepsilon)\left\langle \hat{\mathcal{O}}_{\rm loc}(\varepsilon)\right\rangle \, d\varepsilon  \notag\\
    &\approx\mathcal{I}_{0}-\frac{N}{8}\sum_{\ell\neq\ell'} \lvert\psi_\ell\rvert^2 \lvert\psi_{\ell'}\rvert^2 \cos(\varphi_{\ell,B}-\varphi_{\ell',A})[\ell-\ell']^2 \sigma_{\varphi}^2,
\end{align}
where $\mathcal{I}_0$ is the expected signal in the ideal case. We note that this common-mode error has no effect on ${\ell=\ell'}$ terms in $\mathcal{I}$: thus, the features of the interferometric signal are preserved even with temporal instability, as in a synchronous clock comparison experiment~\cite{oelker2019stability}, so long as the phase offset between the nodes remains small after a coherent evolution.

Similarly, adapting the variance calculation from the Methods section yields
\begin{align}\label{eq:var_epsilon}
    \Delta\mathcal{I}^2&=\int p(\varepsilon)\left\langle \hat{\mathcal{O}}_{\rm loc}(\varepsilon)^2\right\rangle \,d\varepsilon - \mathcal{I}^2 \notag\\
    &= \sum_{\ell,\sigma}\int \frac{\lvert\tilde{\lambda}_{\ell,\sigma}^{\rm (t)}(\varepsilon)\rvert^2}{16}[ (2\ell+1) N^2-2 \ell^2 N \notag\\
    &\qquad\qquad\qquad +\delta_{\ell 2} N(2N-2) ] p(\varepsilon) \,d\varepsilon - \mathcal{I}^2,
\end{align}
where $\tilde{\lambda}_{\ell,\sigma}^{\rm (t)}(\varepsilon)=\bra{\Psi(T)}e^{i\varepsilon\hat{J}_z} \hat{U}_p\otimes e^{i\varepsilon\hat{J}_z}\hat{U}_p \ket{\Psi_{\ell,\sigma}}$. The $\sigma_{\varphi}^2$-order correction can then be derived explicitly, for a given preparation unitary, by computing the integral of $p(\varepsilon)\lvert\tilde{\lambda}_{\ell,\sigma}^{\rm (t)}(\varepsilon)\rvert^2$. Its expression can be further simplified in certain cases, for instance for the basic COW-like excitation, where $\lvert\tilde{\lambda}_{1,+}^{\rm (t)}(\varepsilon)\rvert^2+\lvert\tilde{\lambda}_{1,-}^{\rm (t)}(\varepsilon)\rvert^2=1$ and $\lvert\tilde{\lambda}_{\ell\neq 1,\sigma}^{\rm (t)}\rvert^2=0$. This leads to 
\begin{align}
    \Delta\mathcal{I}^2&\approx\frac{3N^2-2N}{16}-\mathcal{I}_0^2=\Delta\mathcal{I}_0^2,
\end{align}
(since $\mathcal{I}\approx\mathcal{I}_0$ in that case), which is the same expression as without the common-mode phase instability.

We model phase instability in space by adding an random phase $\varepsilon$ (normally distributed with vanishing expectation value, and variance $\sigma_\varphi^2$) to one of the nodes during the decoding and measurement steps. Again based on Ref.~\cite{yan2025highpower}, we estimate that this error is negligible when using the same laser to address both nodes (for inter-node distances on the order of meters), and is approximately one to two orders of magnitude lower than the common-mode instability when using phase-locking (for large distances), i.e., $\sigma_\varphi^2\sim 10^{-6}$ over $25\,\rm ms$. This error corresponds to replacing $\hat{\mathcal{O}}_{\rm loc}$ with $\hat{\mathcal{O}}_{\rm loc}(\varepsilon)=(\mathbb{1}\otimes e^{i\varepsilon \hat{J}_z})\hat{\mathcal{O}}_{\rm loc}(\mathbb{1}\otimes e^{-i\varepsilon \hat{J}_z})$. Using the same calculation as for the temporal instability, this eventually leads to the following leading-order correction to the signal:
\begin{align}
    \mathcal{I}&\approx\mathcal{I}_{0}-\frac{N}{8}\sum_{\ell,\ell'\geq 1} \lvert\psi_\ell\rvert^2 \lvert\psi_{\ell'}\rvert^2 \cos(\varphi_{\ell,B}-\varphi_{\ell',A})\ell^2 \sigma_{\varphi}^2,
\end{align}
i.e., in this case, the noise affects all terms of the interferometric signal.

For the signal uncertainty, we find the same expression as in Eq.~\eqref{eq:var_epsilon}, up to replacing $\tilde{\lambda}_{\ell,\sigma}^{\rm (t)}$ with $\tilde{\lambda}_{\ell,\sigma}^{\rm (s)}(\varepsilon)=\bra{\Psi(T)} (\hat{U}_p\otimes e^{i\varepsilon\hat{J}_z}\hat{U}_p) \ket{\Psi_{\ell,\sigma}}$. Taking again the COW-like case as a basic example, we have also in this case $\lvert\tilde{\lambda}_{1,+}^{\rm (s)}(\varepsilon)\rvert^2+\lvert\tilde{\lambda}_{1,-}^{\rm (s)}(\varepsilon)\rvert^2=1$ and $\lvert\tilde{\lambda}_{\ell\neq 1,\sigma}^{\rm (s)}\rvert^2=0$. Taking into account the correction to $\mathcal{I}$, this leads to
\begin{align}
    \Delta\mathcal{I}^2&\approx\Delta\mathcal{I}_0^2+\mathcal{I}_0^2 \, \ell^2 \sigma_{\varphi}^2,
\end{align}
that is, the signal uncertainty has an extra term of order $\sigma_\varphi ^2$. Quantitatively, for the example of the main text ($N=20$ atoms), this increases the variance by an amount $\sim 400\,\sigma_{\varphi}^2\,\mathcal{I}_0^2$, which is negligible compared to $\Delta \mathcal{I}_0^2 \sim 2\mathcal{I}_0^2$ as long as $\sigma_{\varphi}\ll 10^{-1}$ [in particular, for the typical value given above the correction is $\sim (4\times10^{-4})\,\mathcal{I}_0^2$, so the uncertainty is dominated by quantum fluctuations].

Finally, we note that the signal fluctuations can also be considered in the context of phase estimation sensitivity. Considering a superposition of a single-mass excitation, we can estimate the GR-induced phase $\varphi_{\rm GR}=\varphi_{\ell,B}-\varphi_{\ell,A}$ using the function $\varphi_{\rm GR}(\mathcal{I})=\arccos(4\mathcal{I}/N)$ [see Eq.~(14) of the main text]. Taking a linear approximation for $\varphi_{\rm GR}(\mathcal{I})$, the precision of the phase estimation is given by $\Delta \varphi_{\rm GR}^2 =  \left(\frac{d\varphi_{\rm GR}}{d\mathcal{I}} \right)^2 \Delta \mathcal{I}^2$. Thus, the signal noise discussed above leads to (i) a systematic error in the estimated phase, due to $\mathcal{I}\neq\mathcal{I}_0$, and (ii) an additional uncertainty, given approximately by $4\ell\sigma_{\varphi}\mathcal{I}_0 \,/(N\sin(\varphi_{\rm GR}))$ [which is in general negligible compared to the other term $4\Delta\mathcal{I}_0\,/(N\sin(\varphi_{\rm GR}))$ arising from quantum fluctuations (see Methods)]. 
The phase sensitivity is maximal around $\varphi_{\rm GR}=\pi/2$ (this value is a consequence of our choice of measurement --- in a phase estimation context, the scheme can be straightforwardly adapted to have maximum sensitivity at $\varphi_{\rm GR}=0$); quantitatively, we estimate it to be on the order of $\Delta\varphi_{\rm GR}\sim 0.1$ for the example of the main text ($N=20$ atoms, $\kappa=200$ measurement runs per time point), meaning that the gravity-induced phase can be resolved. 
We emphasize, however, that phase estimation is not the focus of this work, which primary aims to observe gravitationally induced interference effects.

\begin{figure}
    \centering
    \includegraphics[width=\linewidth]{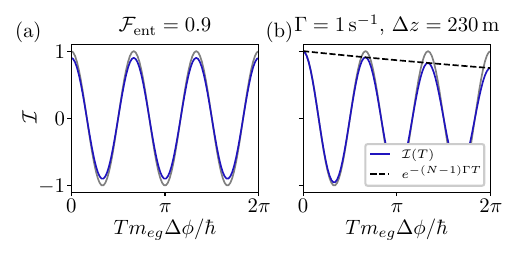}
    \caption{Impact of typical experimental imperfections. (a) Interference pattern with imperfect entanglement distribution (blue line), with Bell state fidelity $\mathcal{F}_{\rm ent}=0.9$ and infidelity dominated by dark counts [Eq.~\eqref{eq:imp_entanglement}]. We consider $N=4$ atoms per node and the preparation circuit $\hat{U}_{\rm DT}$ from the main text. The grey line is the ideal interference pattern. (b) Interference pattern in the presence of spontaneous decay~(blue), for atomic decay rate $\Gamma=\qty{1}{\per\second}$, inter-node height difference $\Delta z=\qty{230}{\meter}$, and clock transition frequency $\omega_{eg}=2\pi\times\qty{411}{\THz}$ (see main text). The signal exhibits exponential decay (dashed black line), in contrast to the Gaussian decay characteristic of gravitational dephasing.}
    \label{fig:imperf}
\end{figure}

\paragraph{Spontaneous decay during interrogation.} Finally, we discuss the impact of spontaneous emission during the Ramsey interrogation step. We do this numerically, introducing spontaneous decay via jump operators $\hat{L}_{n,i}=\sqrt{\Gamma}\ket{g}_{n,i}\bra{e}_{n,i}$ on each atom $i$ of both nodes $n$. Evolution during the Ramsey interrogation time is thus described by the following master equation:
\begin{align}
    \dot{\rho}=-i\left[\hat{H},\hat{\rho}\right]+\sum_{n,i} \left(\hat{L}_{n,i}\hat{\rho}\hat{L}_{n,i}^{\dagger}-\frac{1}{2}\left\{\hat{L}_{n,i}^{\dagger}\hat{L}_{n,i},\hat{\rho}\right\}\right)
\end{align}
[with the GR-corrected Hamiltonian $\hat{H}$ from Eq.~(7) of the main text], where the sum runs over all the atoms of both nodes. With this master equation, we can determine the evolution of the network's state under GR-corrected dynamics, and compute the resulting interference patterns. Supplementary Fig.~\ref{fig:imperf} shows such a result for typical experimental parameters discussed in the main text for the trapped-ion case, and for the $\hat{U}_{\rm DT}$ exact preparation circuit. We observe a decay of the interference pattern with rate $(N-1)\Gamma$. This decay is exponential, and thus distinct from the Gaussian decay arising from gravitational dephasing. We note that beyond spontaneous emission, other platform-dependent decoherence channels can adversely affect our interference signal, including single-particle and collective dephasing noise~\cite{kraft2025boundederrorquantumsimulationhamiltonian}. 

\section{Alternative measurement schemes}\label{sec:alt_measurements}

We now discuss alternatives to the local measurement scheme presented in the main text, specifically: (i) A non-local measurement scheme relying on reading out the ensemble memories, which more closely reflects atom- and atom-clock-interferometer setups; (ii) A local measurement scheme based on individual readout of the qubits in the nodes, adapted for near-term trapped-ion implementations.

\paragraph{Non-local scheme.} In contrast to the local measurement scheme presented in the main text, this scheme does not require applying a decoding unitary (i.e., $\hat{U}_m=\mathbb{1}$). After the Ramsey interrogation, the scheme consists in first reading out each node's quantum memory into an photon channel (with the excitation number becoming the photon number), and then combining the two output beams in a beam splitter. The photon numbers $\hat{N}_1$ and $\hat{N}_2$ in the two output modes then contain information on the common state of nodes $A$ and~$B$: specifically, the parity of the photon number $\hat{N}_2$ reveals information on the phases $\varphi_{\ell,n}$ as (taking $\varphi_0=0$ for simplicity)
\begin{align}\label{eq:I_nonlocal}
    \mathcal{I}(T) = \left\langle (-1)^{\hat{N}_2} \right\rangle=\sum_{\ell\geq1}\lvert\psi_\ell\rvert^2\cos(\varphi_{\ell,B}-\varphi_{\ell,A})
\end{align} 
which reproduces the atom-interferometer result of Eq.~(21) of the Methods section.

This can be seen as follows. The resulting state after excitation-to-photon readout can be written as $\ket{\Psi(T)}$ from Eq.~(13) of the main text, although with the $\ket{\ell}$ basis now denoting photon number eigenstates. We then combine the outputs from nodes $A$ and $B$ into a beam splitter (with output ports denoted as $1$ and $2$), which we take to be described by the transfer matrix
\begin{align}
    U_{\rm BS}=\frac{1}{\sqrt{2}}
    \begin{pmatrix}
        1 & 1\\
        1 & -1
    \end{pmatrix},
\end{align}
such that the field operators evolve according to 
\begin{align}
    \begin{pmatrix}
        \hat{a}^{\dagger}_1\\
        \hat{a}^{\dagger}_2
    \end{pmatrix}
    =U_{\rm BS}
    \begin{pmatrix}
        \hat{a}^{\dagger}_A\\
        \hat{a}^{\dagger}_B
    \end{pmatrix}.
\end{align}
With $\ket{\Psi(T)}$ as input state, the beam splitter's output is
\begin{align}
    \ket{\Psi_{\rm out}}
    &=\sum_{\ell\geq1}\frac{\psi_\ell}{\sqrt{\ell!}\,(\sqrt{2})^{\ell+1}}\sum_{\ell'\leq \ell}\binom{\ell}{\ell'}(e^{-i\varphi_{\ell,A}}+{(-1)^{\ell'}} e^{-i\varphi_{\ell,B}})\notag\\
    &\qquad\qquad\qquad\qquad(\hat{a}^{\dagger}_2)^{\ell'}(\hat{a}^{\dagger}_1)^{\ell-\ell'}\ket{\rm vac},
\end{align}
where $\hat{a}^{\dagger}_1$ and $\hat{a}^{\dagger}_2$ are photon creation operators in output ports $1$ and $2$. Now, measuring the parity of the photon number in output port $2$ [i.e., the observable $\hat{\mathcal O}_{\rm nl}\equiv(-1)^{\hat{N}_2}$] gives:
\begin{align}
    P(\pm 1)&=\sum_{\ell' \;{\rm even/odd}}\sum_{\ell\geq \ell'}\frac{\lvert\psi_\ell\rvert^2}{2^{\ell+1}}\binom{\ell}{\ell'}\lvert e^{-i\varphi_{\ell,A}}\pm e^{-i\varphi_{\ell,B}}\rvert^2\\
    &=\sum_{\ell\geq 1}\frac{\lvert\psi_\ell\rvert^2}{2}[1\pm\cos(\varphi_{\ell,B}-\varphi_{\ell,A})],
\end{align}
Giving the expression shown in Eq.~\eqref{eq:I_nonlocal} for the expectation value $\mathcal{I}(T)$. We note that this result coincides with the result of Eq.~(21) of the Methods section.

This scheme is conceptually interesting in that it does not require the application of a decoding unitary, and perfectly mirrors the interference signals obtained in matter-wave interferometry setups. However, it involves a truly non-local entangling operation between the two nodes (via the beam splitter), and requires faithful conversion of excitation to photons and a number-resolving photodetector; this makes a large-$N$ implementation of this scheme challenging with currently available technologies.

\paragraph{Individual Pauli measurements} In the scenario discussed above (Sec.~\ref{sec:alt_states}), in which the initial seed state is a Bell state on two atoms only, the same interference signal $\mathcal{I}(T)$ as in the main text can be obtained by replacing the collective $\hat{J}_x$ measurements with a $\hat{\sigma}_x$ Pauli measurement on the first qubit of each node. The derivation of the interference signal and of its uncertainty follows the same steps as in the Methods section, with the measurement observable replaced with $\hat{\mathcal{O}}_{\rm loc, Pauli}=(\hat{U}_p\otimes \hat{U}_p)(\hat{\sigma}_x^{(a)}\otimes\hat{\sigma}_x^{(b)})(\hat{U}_p^{\dagger}\otimes\hat{U}_p^{\dagger})$, with $a$ and $b$ denoting the first atom of each ensemble. Following the steps, we obtain the expectation value
\begin{align}
    \mathcal{I}(T)&=\bra{\Psi(T)}\hat{\mathcal{O}}_{\rm loc, Pauli}\ket{\Psi(T)}\\
    &=\sum_{\ell,\ell'\geq 1}\lvert\psi_\ell\rvert^2 \lvert\psi_{\ell'}\rvert^2 \cos(\varphi_{\ell,B}-\varphi_{\ell',A}),
\end{align}
which is the result of the main text up to a prefactor, and the measurement variance
\begin{align}
    \Delta\mathcal{I}^2(T)&=\left\langle \hat{\mathcal{O}}_{\rm loc, Pauli} ^2\right\rangle - \left\langle \hat{\mathcal{O}}_{\rm loc, Pauli} \right\rangle^2\\
    &=1-\mathcal{I}(T)^2.
\end{align}
In this case, the interferometric signal and its uncertainty both do not scale with $N$.